\documentclass[twoside,twocolumn,9pt]{article}
\usepackage{extsizes}
\usepackage[super,sort&compress,comma]{natbib} 
\usepackage[version=3]{mhchem}
\usepackage[left=1.5cm, right=1.5cm, top=1.785cm, bottom=2.0cm]{geometry}
\usepackage{balance}
\usepackage{mathptmx}
\usepackage{sectsty}
\usepackage{graphicx} 
\usepackage{lastpage}
\usepackage[format=plain,justification=justified,singlelinecheck=false,font={stretch=1.125,small,sf},labelfont=bf,labelsep=space]{caption}
\usepackage{float}
\usepackage{fancyhdr}
\usepackage{fnpos}
\usepackage[english]{babel}
\addto{\captionsenglish}{%
  
}
\usepackage{array}
\usepackage{droidsans}
\usepackage{charter}
\usepackage[T1]{fontenc}
\usepackage[usenames,dvipsnames]{xcolor}
\usepackage{setspace}
\usepackage[compact]{titlesec}
\usepackage{hyperref}
\usepackage{braket}
\usepackage{amsmath}
\usepackage{amssymb}
\usepackage{tikz}
\usetikzlibrary{positioning}
\usetikzlibrary{calc}
\usetikzlibrary{decorations.markings}
\usetikzlibrary {arrows.meta}
\usetikzlibrary{snakes}
\usepackage[sep=0.4em, offset=1.4em]{simpler-wick} 

\definecolor{cream}{RGB}{222,217,201}

\begin{document}

\pagestyle{fancy}
\thispagestyle{plain}
\fancypagestyle{plain}{
%%%HEADER%%%
\renewcommand{\headrulewidth}{0pt}
}
%%%END OF HEADER%%%

%%%PAGE SETUP - Please do not change any commands within this section%%%
\makeFNbottom
\makeatletter
\renewcommand\LARGE{\@setfontsize\LARGE{15pt}{17}}
\renewcommand\Large{\@setfontsize\Large{12pt}{14}}
\renewcommand\large{\@setfontsize\large{10pt}{12}}
\renewcommand\footnotesize{\@setfontsize\footnotesize{7pt}{10}}
\makeatother

\renewcommand{\thefootnote}{\fnsymbol{footnote}}
\renewcommand\footnoterule{\vspace*{1pt}% 
\color{cream}\hrule width 3.5in height 0.4pt \color{black}\vspace*{5pt}} 
\setcounter{secnumdepth}{5}

\makeatletter 
\renewcommand\@biblabel[1]{#1}            
\renewcommand\@makefntext[1]% 
{\noindent\makebox[0pt][r]{\@thefnmark\,}#1}
\makeatother 
\renewcommand{\figurename}{\small{Fig.}~}
\sectionfont{\sffamily\Large}
\subsectionfont{\normalsize}
\subsubsectionfont{\bf}
\setstretch{1.125} %In particular, please do not alter this line.
\setlength{\skip\footins}{0.8cm}
\setlength{\footnotesep}{0.25cm}
\setlength{\jot}{10pt}
\titlespacing*{\section}{0pt}{4pt}{4pt}
\titlespacing*{\subsection}{0pt}{15pt}{1pt}
%%%END OF PAGE SETUP%%%

%%%FOOTER%%%
\fancyfoot{}
\fancyfoot[LO,RE]{\vspace{-7.1pt}\includegraphics[height=9pt]{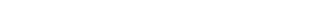}}
\fancyfoot[CO]{\vspace{-7.1pt}\hspace{11.9cm}\includegraphics{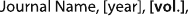}}
\fancyfoot[CE]{\vspace{-7.2pt}\hspace{-13.2cm}\includegraphics{head_foot/RF}}
\fancyfoot[RO]{\footnotesize{\sffamily{1--\pageref{LastPage} ~\textbar  \hspace{2pt}\thepage}}}
\fancyfoot[LE]{\footnotesize{\sffamily{\thepage~\textbar\hspace{4.65cm} 1--\pageref{LastPage}}}}
\fancyhead{}
\renewcommand{\headrulewidth}{0pt} 
\renewcommand{\footrulewidth}{0pt}
\setlength{\arrayrulewidth}{1pt}
\setlength{\columnsep}{6.5mm}
\setlength\bibsep{1pt}
%%%END OF FOOTER%%%

%%%FIGURE SETUP - please do not change any commands within this section%%%
\makeatletter 
\newlength{\figrulesep} 
\setlength{\figrulesep}{0.5\textfloatsep} 

\newcommand{\topfigrule}{\vspace*{-1pt}% 
\noindent{\color{cream}\rule[-\figrulesep]{\columnwidth}{1.5pt}} }

\newcommand{\botfigrule}{\vspace*{-2pt}% 
\noindent{\color{cream}\rule[\figrulesep]{\columnwidth}{1.5pt}} }

\newcommand{\dblfigrule}{\vspace*{-1pt}% 
\noindent{\color{cream}\rule[-\figrulesep]{\textwidth}{1.5pt}} }

\makeatother
%%%END OF FIGURE SETUP%%%

%%%TITLE, AUTHORS AND ABSTRACT%%%
\twocolumn[
  \begin{@twocolumnfalse}
{\includegraphics[height=30pt]{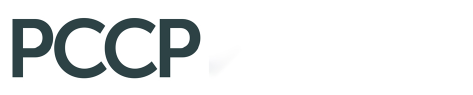}\hfill\raisebox{0pt}[0pt][0pt]{\includegraphics[height=55pt]{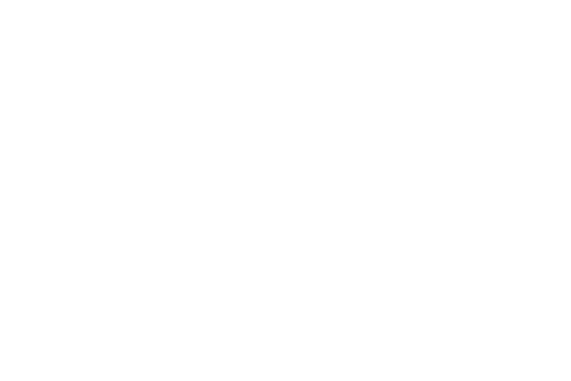}}\\[1ex]
\includegraphics[width=18.5cm]{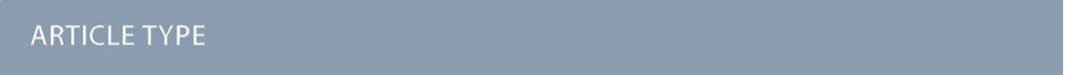}}\par
\vspace{1em}
\sffamily
\begin{tabular}{m{4.5cm} p{13.5cm} }

\includegraphics{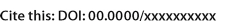} & \noindent\LARGE{\textbf{Multi-reference coupled cluster theory using the normal ordered exponential ansatz}} \\%Article title goes here instead of the text "This is the title"
\vspace{0.3cm} & \vspace{0.3cm} \\

 & \noindent\large{Alexander Gunasekera,\textit{$^{1}$}$^{\ast}$ Nicholas Lee,\textit{$^{1}$}$^{\ast}$ and David P. Tew\textit{$^{1}$}$^{\ast}$} \\

\includegraphics{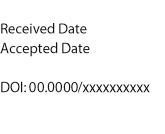} & \noindent\normalsize{    Properly spin-adapted coupled-cluster theory for general open-shell configurations remains an active area of research in electronic structure theory. In this contribution we examine Lindgren’s normal-ordered exponential ansatz to correlate specific spin states using spin-free excitation operators, with the aid of automatic equation generation software. We present an intermediately normalised and size-extensive reformulation of the unlinked working equations, and analyse the performance of the method with single and double excitations for simple molecular systems in terms of accuracy and size-consistency.} \\

\end{tabular}

 \end{@twocolumnfalse} \vspace{0.6cm}

  ]
%%%END OF TITLE, AUTHORS AND ABSTRACT%%%

%%%FONT SETUP - please do not change any commands within this section
\renewcommand*\rmdefault{bch}\normalfont\upshape
\rmfamily
\section*{}
\vspace{-1cm}

%%%FOOTNOTES%%%

\footnotetext{\textit{$^{1}$~Physical and Theoretical Chemistry Laboratory, University of Oxford, South Parks Road, Oxford, OX1 3QZ, UK}}
\footnotetext{\textit{$^{\ast}$~Corresponding author email: alexander.gunasekera@chem.ox.ac.uk}}
\footnotetext{\textit{$^{\ast}$~Corresponding author email: nicholas.lee@chem.ox.ac.uk}}
\footnotetext{\textit{$^{\ast}$~Corresponding author email: david.tew@chem.ox.ac.uk}}

%%%END OF FOOTNOTES%%%

%%%MAIN TEXT%%%%

\section{Introduction}
Coupled cluster (CC) theory based on a single Slater-determinant reference wavefunction is firmly established and is widely used in high-accuracy electronic structure calculations\cite{bartlett_coupled-cluster_2007}. 
%Coupled cluster is preferable to perturbation theory since it can capture dynamic correlation and orbital relaxation and does not rely on a zeroth order Hamiltonian. 
Many chemical systems, including diradicals, transition metals, and molecules at non-equilibrium geometries, display open-shell configurations for which multi-determinant reference states are required.
%When symmetry cannot be exploited to define a single reference function, true 
The analogous multi-reference coupled-cluster (MRCC) theory for correlating open-shell systems remains an area of active research due to challenges arising from the complexity of defining wave operators and working equations for multi-determinant reference spaces\cite{kohn_state-specific_2012, lyakh_multireference_2011, evangelista_perspective_2018}. 
%No consensus method has yet emerged.
%Effective Hamiltonian formulations for correlating the multiple reference determinants allow for the simultaneous computation of ground and excitated state energies\cite{?????}. 
Our work concerns the generalisation of closed-shell coupled-cluster theory to  arbitrary open-shell states while retaining full spin-adaption, through a state-specific formulation. State-specific MRCC \textit{ansätze} can be broadly classified into two categories:
those that define a different wave operator for each reference configuration are known as Jeziorski--Monkhorst (JM) \textit{ansätze}\cite{jeziorski_coupled-cluster_1981};
the other type, in which a single wave operator is applied to a linear combination of the reference functions, are known as internally contracted \textit{ansätze}\cite{banerjee_coupled-cluster_1981, evangelista_sequential_2012, hanauer_pilot_2011}. These methods are multi-reference because they include linear or non-linear parameters that are optimised separately for each contributing reference determinant. 
Owing to the limitations of the single Slater determinant picture, open-shell systems are often misidentified as multi-reference. 
Multi-reference treatments are necessary when competing configurations are involved in bonding or when targeting excited state spectra, but are not necessary\cite{martidafcik2024spincoupled} for states that are correctly represented by a 
single open-shell configuration state function (CSF), which are eigenfunctions of the spin operators. 
Many open-shell systems, such as organic radicals and transition metal spin states may be treated using single-reference open-shell coupled-cluster theory.
In these cases, it is not necessary to perform a full CAS calculation, whose cost formally scales exponentially with the number of open-shell electrons.
In this work, we explore fully spin-adapted coupled-cluster theory for correlating single open-shell CSF reference states. 
Our \textit{ansatz} takes the form of the internally contracted theories, but where the coefficients of the multi-determinant reference state are pre-determined by spin and spatial symmetry constraints.
Full spin-adaptation of is achieved through the use of spin-free excitation operators in our cluster operator, in a manner similar to unitary group approaches\cite{xiangzhu_multiconfigurational_1993, xiangzhu_automation_1994, xiangzhu_unitary_1997, xiangzhu_unitary_1998, xiangzhu_unitary-group-based_1998}.
%This is similar in  to the Unitary Group Approach (UGA) methods\cite{xiangzhu_multiconfigurational_1993, xiangzhu_automation_1994, xiangzhu_unitary_1997, xiangzhu_unitary_1998, xiangzhu_unitary-group-based_1998}, where cluster operators are adapted to the chain of unitary subgroups enforcing orbital invariance within the separate core, active, and virtual spaces.

% Unitary Group Approaches (UGAs) to the spin-adapted CC for single CSFs were introduced by Li and Paldus in a series of papers in the 1990s.
% In their work, a full exponential is used in analogy with the closed shell coupled cluster, and the cluster operators are further adapted to the chain of unitary subgroups enforcing orbital invariance within the separate core, active, and virtual spaces.

Since excitations involving singly-occupied orbitals do not commute, we adopt Lindgren’s normal-ordered exponential (NOE) form of the coupled-cluster wave operator\cite{lindgren_coupled-cluster_1978}.
This choice ensures that there are no contractions among cluster operators and thus the working equations terminate at finite order in the cluster amplitudes. 
Single CSF references are invariant with respect to rotations among the doubly-occupied orbitals, but not in general among the open-shell (active) orbitals. We include purely active-to-active excitations in our cluster operator, since these are required to fully allow for correlation-induced orbital relaxation of the CSF reference. 
%This feature distinguishes our work from similar coupled-cluster theories.
%The problem of non-commuting cluster operators may be alleviated by the normal-ordered exponential coupled cluster (NOE-CC) ansatz proposed by Lindgren in 1978.
The normal-ordered \textit{ansatz} allows the inclusion of these otherwise problematic excitations. It is our opinion that this \textit{ansatz}, with its origins in the factorisation theorem, best aligns with the physical motivation for coupled cluster as modelling independent excitation events.

The NOE \textit{ansatz} has previously been considered in the context of Fock-Space MRCC\cite{haque_application_1984, stolarczyk_coupled-cluster_1985, stolarczyk_coupled-cluster_1985-1, stolarczyk_coupled-cluster_1988, stolarczyk_coupled-cluster_1988-1, kaldor_fock_1991} and the Similarity Transformed Equation-of-Motion Coupled Cluster\cite{nooijen_general_1996, nooijen_manybody_1996, nooijen_new_1997, nooijen_similarity_1997, nooijen_similarity_1997-1, nooijen_communication_2014} (ST-EOM-CC) approaches.
In these two methods, normal-ordering of the exponential allows for a partial decoupling of the different valence sectors, simplifying the working equations. Mukherjee and co-workers have explored the NOE \textit{ansatz} with unitary group adapted cluster operators for state-universal and state-specific MRCC theories, using a JM \textit{ansatz} in an effective Hamiltonian formulation
\cite{jana_compact_2002, datta_compact_2008, datta_explicitly_2009,maitra_unitary_2012,Sinha_ugassmrcc_2012,Sen_uga-sumrcc_2012}.
Very recently, they succeeded in replacing the sufficiency conditions used in their earlier work with a rigorous effective Hamiltonian formulation\cite{sen_inclusion_2018,chakravarti_reappraisal_2021,chakravarti_ugassmrccsd_2023}.
The single-reference limit of Mukherjee's approach\cite{sen_inclusion_2018} reduces to the same 
NOE \textit{ansatz} as used in this work. 

%A variation on the NOE \textit{ansatz}, where only contractions through spectator orbitals are allowed, have also been considered by Mukherjee and co-workers in a Combinatoric open-shell coupled cluster theory (COS-CC\cite{jana_compact_2002, datta_compact_2008, datta_explicitly_2009}).
%Subsequently, Mukherjee \textit{et al.} have considered the normal ordered exponential with a JM \textit{ansatz}, and shown that it reproduces the more complicated COS-CC theory, despite the neglect of contractions through spectating orbitals\cite{jana_compact_2002, datta_compact_2008, datta_explicitly_2009}.

In this contribution we present an alternative and much simpler formulation of the working equations for NOECC theory that permits systematic approximation. We generate our working equations via an intermediately normalised unlinked CC formalism in such a way as to preserve size extensivity.
We demonstrate through analysis and numerical testing that our methods are rigorously spin-adpated, and that the errors in size-extensivity and size-consistency for homolytic bond fission introduced by truncation of our theory may be systematically removed at a finite order in the cluster operator.
Our approach also differs from previous work in that we formally allow purely active-to-active excitations in order to treat active orbital relaxation, and that we do not insist on spin-completeness of the excited state projection manifold.
Due to the complexity of obtaining spin-free working equations with the NOE \textit{ansatz}, we have implemented our own domain-specific Wick contraction engine and automated equation generator, using similar ideas to those of Hermann and Hanrath\cite{hermann_generation_2020, hermann_analysis_2021, hermann_correctly_2022}.

\section{Theory}

\subsection{Preliminaries}
We employ the following convention for orbital labelling: $i,j,k,\dots$ for core (doubly-occupied) orbitals, $a,b,c,\dots$ for virtual (unoccupied) orbitals, $t,u,v,\dots$ for active orbitals, and $p,q,r,\dots$ for general orbitals. The core, active, and virtual spaces will be denoted $\mathbb{C}$, $\mathbb{A}$, and $\mathbb{V}$ respectively.
We express creation and annihilation operators using the tensor notation $\hat{a}^{p\sigma} = \hat{a}_{p\sigma}^{\dagger}$ and $\hat{a}_{p\sigma} = \hat{a}_{p\sigma}$, and define spin-free unitary group generators
\begin{align}
     \hat{E}_{q}^{p}&=\sum_{\sigma}\hat{a}^{p\sigma}\hat{a}_{q\sigma}\\
    \hat{E}_{rs}^{pq}&= \sum_{\sigma\tau}\hat{a}^{p\sigma}\hat{a}^{q\tau}\hat{a}_{s\tau}\hat{a}_{r\sigma}
\end{align}
and so on for higher-body operators.
Our cluster operator is a sum of n-body spin-free excitation operators $\hat{T}_n$ that each commute with the total spin squared operator $\hat{S}^{2}$.
\begin{equation}
\hat{T}=\hat{T}_{1}+\hat{T}_{2}+\dots=\sum_{\substack{p \in \mathbb{A} \cup \mathbb{V} \\ q \in \mathbb{C} \cup \mathbb{A}}}t_{p}^{q}\,\left\{\hat{E}_{q}^{p}\right\}+\frac{1}{2}\sum_{\substack{p,q \in \mathbb{A} \cup \mathbb{V} \\ r,s \in \mathbb{C} \cup \mathbb{A}}}t_{pq}^{rs}\,\left\{\hat{E}_{rs}^{pq}\right\} +\dots
\end{equation}
where the braces $\left\{\right\}$ denote normal ordering with respect to the closed shell vacuum corresponding to doubly occupying the core orbitals and leaving the active and virtual orbitals vacant.

We consider the spin-free molecular electronic Hamiltonian $\hat{H}$ in normal order\cite{kutzelnigg_normal_1997} with respect to the same vacuum:
\begin{align}
    \hat{H} &= \sum_{pq} f_{p}^{q} \left\{\hat{E}_{q}^{p}\right\} + \frac{1}{2} \sum_{pqrs} g_{pq}^{rs}\left\{\hat{E}_{rs}^{pq}\right\}
\end{align}
with $h_{p}^{q} = \braket{p | \hat{h} | q}$ and $g_{pq}^{rs} = \braket{pq | r_{12}^{-1} | rs} = (pr|r_{12}^{-1} |qs)$, and the core Fock matrix is defined by $f_{p}^{q} = h_{p}^{q} + \sum_{i}\left(2g_{pi}^{qi}-g_{pi}^{iq}\right)$. 

The open-shell part of the reference is a CSF of $N$ electrons in $N$ orbitals with total spin $S$ and projected spin $M$, expressed through a linear combination of creation operator strings acting upon a closed shell vacuum, $\Ket{\Phi_{0}}=\Ket{N,S,M;\mathbf{t}}= \hat{O}^{S,M}_{N}(\mathbf{t}) \ket{0}$\cite{helgaker_modern_2013}. The genealogical coupling vector $\mathbf{t}$ specifies a particular Gel'fand--Tsetlin state in cases of spin degeneracy when there are more than two open-shell electrons.
%The creation operator $\hat{O}^{S,M}_{N}(\mathbf{t})$ for a particular configuration is specified by a genealogical coupling vector $\mathbf{t}$ to give spin $S$ and spin projection $M$.
The creation operator $\hat{O}^{S,M}_{N}(\mathbf{t})$ is built up by recursively applying the definition
\begin{equation}
\hat{O}^{S,M}_{N}(\mathbf{t}) = C^{S,M}_{t_{N}, \frac{1}{2}} \hat{O}^{S-t_{N},M-\frac{1}{2}}_{N-1}(\mathbf{t}) \hat{a}^{N \alpha} + C^{S,M}_{t_{N}, -\frac{1}{2}} \hat{O}^{S-t_{N},M+\frac{1}{2}}_{N-1}(\mathbf{t}) \hat{a}^{N \beta}
\end{equation}
This recursive construction is known as a genealogical coupling scheme as it depends on the history, specified by $\mathbf{t}$, of the configurations as the angular momentum of each electron is added in turn.
Each component $t_{N}$ denotes whether each electron increases the total spin ($t_{N}=+\frac{1}{2}$) or decreases it ($t_{N}=-\frac{1}{2}$) when added ($\hat{a}^{N \alpha/\beta}$) in the $N$-th orbital to the previous $\left(N-1\right)$-electron CSF.
The Clebsch-Gordan coefficients $C^{S,M}_{t_{N}, \pm \frac{1}{2}}$ for addition of a single electron simplify to $C^{S,M}_{+\frac{1}{2}, \pm \frac{1}{2}}=\sqrt{\frac{S\pm M}{2S}}$ and $C^{S,M}_{-\frac{1}{2}, \pm \frac{1}{2}}=\mp\sqrt{\frac{S\mp M + 1}{2S+2}}$.
For example, an open-shell singlet in orbitals $p$ and $q$ is formed from the closed shell vacuum $\ket{0}$ by the operator
\begin{equation}
    \hat{O}^{0,0}_{2} = \frac{1}{\sqrt{2}} \Big( \hat{a}^{p \alpha} \hat{a}^{q \beta} - \hat{a}^{p \beta} \hat{a}^{q \alpha} \Big)
\label{eqn:csf-singlet}
\end{equation}
and the $M=0$ component of a triplet by
\begin{equation}
\hat{O}^{1,0}_{2} = \frac{1}{\sqrt{2}} \Big(  \hat{a}^{p \alpha} \hat{a}^{q \beta} + \hat{a}^{p \beta} \hat{a}^{q \alpha} \Big)
\label{eqn:csf-triplet}
\end{equation}

\subsection{Normal Ordered Exponential Coupled Cluster}
The traditional coupled cluster wavefunction for a cluster operator containing terms that mutually commute is written as an exponential wave operator applied to the reference determinant
\begin{equation}
    \Ket{\Psi}=\mathrm{e}^{\hat{T}}\Ket{\Phi_{0}}
\end{equation}
Projecting the Schr\"odinger equation for this \textit{ansatz} onto a manifold of excited states $\Ket{\Phi_{I}}$ yields the coupled cluster equations in their energy-dependent, ``unlinked" formulation
\begin{align}
    E &=\braket{\Phi_{0}|\hat{H} \mathrm{e}^{\hat{T}} |\Phi_{0}} \\
    0=R_{I}&=\braket{\Phi_{I}|(\hat{H} - E)\mathrm{e}^{\hat{T}}|\Phi_{0}}
\end{align}
It is common to use the alternative ``linked" formalism, which pre-multiplies the Schr\"odinger equation by the inverse wave operator $\mathrm{e}^{-\hat{T}}$ to give an effective Hamiltonian $\Tilde{H}=\mathrm{e}^{-\hat{T}}\hat{H} \mathrm{e}^{\hat{T}}=\left(\hat{H} \mathrm{e}^{\hat{T}}\right)_{\mathrm{C}}$ that contains only linked diagrams, with the CC equations
\begin{align}
    E &=\braket{\Phi_{0}|\Tilde{H} |\Phi_{0}} \\
    0=R_{I}&=\braket{\Phi_{I}|\Tilde{H}|\Phi_{0}}
\end{align}
retaining size-extensivity for each term in the cluster operator, and terminating at fourth order in the amplitudes, as shown by the Baker--Campbell--Hausdorff (BCH) expansion for $\Tilde{H}$ in nested commutators.
%$\Tilde{H}=\hat{H}+\left[\hat{H},\hat{T}\right]+\frac{1}{2!}\left[\left[\hat{H},\hat{T},\hat{T}\right]\right]+...$
The single-reference closed-shell and spin-unrestricted open-shell CC theories use cluster operators that mutually commute, and so are almost always implemented in the linked formalism.
The appropriate generalisation for an open-shell CSF reference state $\Ket{\Phi_{0}}$ is Lindgren's normal-ordered exponential coupled-cluster (NOECC) wave operator
\begin{equation}
%    \Ket{\Psi_{\mathrm{NOECC}}}=\left\{\mathrm{e}^{\hat{T}}\right\}\Ket{\Phi_{0}}
    \Ket{\Psi}=\{\mathrm{e}^{\hat{T}}\}\Ket{\Phi_{0}}
\end{equation} 
where braces indicate normal-ordering with respect to the closed-shell vacuum. 
The normal-ordered exponential form excludes contractions between cluster operators and properly parametrises independent correlation processes according to the factorisation theorem\cite{lindgren_coupled-cluster_1978}. Without normal-ordering, contractions among active-to-active excitations in the cluster operator, such as $\hat{T}_{t}^{u}$ and $\hat{T}_{tx}^{uy}$, would give rise to non-terminating working equations. Normal-ordering ensures that the energy and amplitude equations terminate at finite order in the cluster operators. The order at which the equations terminate is $4+\min(n,2N_{r})$, where $n$ is the number of active orbitals and $N_r$ is the maximum rank of the excitation operators. When applied to a closed-shell reference, $\{\mathrm{e}^{\hat{T}}\}  = \mathrm{e}^{\hat{T}}$ and NOECC is equivalent to the standard coupled-cluster ansatz.
Internally contracted MRCC can be formulated with linked equations because active-to-active excitations are excluded.
Using a linked formalism for NOECC theory in an analogous way to the standard theory would require knowledge of the inverse wave operator $\{ e^{\hat{T}} \}^{-1}$, which is non-trivial\cite{nooijen_manybody_1996}.
Several attempts have been made to construct linked equations equivalent to a similarity transformed Hamiltonian by other means. The work by Mukherjee \textit{et al.} uses the Bloch formalism
\begin{align}
\hat{H}\{ \mathrm{e}^{\hat{T}}  \}|\Phi_{0}\rangle = \{ \mathrm{e}^{\hat{T}}  \} \hat{H}_\text{eff} |\Phi_{0}\rangle
\end{align}
where $\hat{H}_\text{eff}$ is an effective Hamiltonian in the reference space $|\Phi_{0}\rangle$ with the same eigenenergy as the target space $\{ \mathrm{e}^{\hat{T}}\} |\Phi_{0}\rangle$. By applying Wick's theorem to factor out $\{ \mathrm{e}^{\hat{T}}\}$ they obtain a rigorous expression for the effective Hamiltonian as an infinite series, along with a hierarchy of finite order approximations of increasing accuracy\cite{sen_inclusion_2018, chakravarti_reappraisal_2021}. Intermediate normalisation $\hat P \{ \mathrm{e}^{\hat{T}} \} \hat P = \hat P$ is not imposed since it is in general incompatible with size-extensivity, which requires $(1-\hat P) \hat H_\text{eff}\hat P=0$, where $\hat P$ projects onto the reference space\cite{lindgren_linked-diagram_1985,lindgren_connectivity_1987}. 
Our approach is much simpler. Recognising that for the single-reference case intermediate normalisation can be imposed, we use the unlinked form.

The unlinked energy and amplitude equations are given by 
\begin{align}
\label{Theory:EnergyEq}
%    E_{\mathrm{NOECC}}=\braket{\Phi_{0}|\hat{H}\left\{ \mathrm{e}^{\hat{T}} \right\}|\Phi_{0}} 
    E &=\braket{\Phi_{0}|\hat{H}\{ \mathrm{e}^{\hat{T}} \}|\Phi_{0}} \\
    R_{I}&=\braket{\Phi_{I}|(\hat{H} - E)\{\mathrm{e}^{\hat{T}}\}|\Phi_{0}}
\end{align}
where the first-order interacting space is $\langle \Phi_I \vert = \langle \Phi_0 \vert \hat E_{I}^{\dag}$, with $I$ standing for one of the cluster excitations.
The unlinked residual equations can be rewritten as
\begin{align}
%    R_{i}&=\braket{\Phi_{i}|(\hat{H} - E_{\mathrm{NOECC}})\left\{\mathrm{e}^{\hat{T}}\right\}|\Phi_{0}}\\
    R_{I}&= \braket{\Phi_{I}|\hat{H}\{\mathrm{e}^{\hat{T}}\}|\Phi_{0}} -\braket{\Phi_{I}|\{\mathrm{e}^{\hat{T}}\}|\Phi_{0}} E\\
    %{\mathrm{e}^{\hat{T}}\right\}|\Phi_{0}} E_{\mathrm{NOECC}}\\
    &=\braket{\Phi_{I}|\hat{H}\{\mathrm{e}^{\hat{T}}\}|\Phi_{0}} - \braket{\Phi_{I}|\{\mathrm{e}^{\hat{T}}\}|\Phi_{0}} \braket{\Phi_{0}|\hat{H}\{\mathrm{e}^{\hat{T}}\}|\Phi_{0}}
\label{Theory:AmplitudeEq}
\end{align}
A proof that Eq. \ref{Theory:AmplitudeEq} is equivalent to a connected form of the residual equations and that the formulation is therefore size extensive is provided in Appendix \ref{A-Connectedness}. The equations terminate, but at a high order in $\hat T$ of $4+\min(n,2N_{r})$. In this contribution, we study truncated equations. 
Naively truncating the exponential in the unlinked formalism to obtain a lower-cost approximation would result in non-size extensive equations. 
%but since the equations are connected at each order in $\hat T$, we can 
However, the size-extensivity error can be kept small by retaining all terms in Eq. \ref{Theory:AmplitudeEq} up to a given power in the amplitudes (see Appendix \ref{A-Connectedness}). 

Specifically, the linearised equations (l-NOECC) are:
\begin{align}
    E&=\braket{\Phi_{0}|\hat{H}\{1 + \hat{T}\}|\Phi_{0}} \\
    R_{I}&=\braket{\Phi_{I}|\hat{H}\{1 + \hat{T}\}|\Phi_{0}} - \braket{\Phi_{I}|\{\hat{T}\}|\Phi_{0}} \braket{\Phi_{0}|\hat{H}|\Phi_{0}}
\end{align}
and the quadratic equations (q-NOECC) are:
\begin{align}
    E=&\braket{\Phi_{0}|\hat{H}\{1 + \hat{T} + \frac{1}{2} \hat{T}^{2} \}|\Phi_{0}} \\
    R_{I}=& \braket{\Phi_{I}|\hat{H}\{1 + \hat{T} + \frac{1}{2} \hat{T}^{2} \}|\Phi_{0}} - \braket{\Phi_{I}|\{\hat{T} \}|\Phi_{0}} \nonumber \braket{\Phi_{0}|\hat{H}\{1 + \hat{T}\}|\Phi_{0}} \\
    &- \braket{\Phi_{I}|\{\frac{1}{2} \hat{T}^{2} \}|\Phi_{0}} \braket{\Phi_{0}|\hat{H}|\Phi_{0}}
\end{align}
Since standard CCD only includes terms up to quadratic in the amplitudes, our truncated q-NOECCD method exactly replicates closed-shell CCD, and q-NOECCSD differs from CCSD only by neglecting third- and fourth-order terms involving singles amplitudes and the failure to cancel the disconnected $\braket{\Phi_{I}|\{\hat{T} \}|\Phi_{K}} \nonumber \braket{\Phi_{K}|\hat{H}\{\hat{T}\}|\Phi_{0}}$ term in the residual equation, where $\ket{\Phi_{K}}$ is a singly excited state and $\ket{\Phi_{I}}$ is doubly excited. 

Our unlinked formulation explicitly imposes intermediate normalisation. It therefore allows for the possibility to include purely active-to-active excitations in the cluster operator that would formally break intermediate normalisation, since the amplitudes that would violate intermediate normalisation are necessarily zero at convergence. 

\subsection{The Cluster Operator}
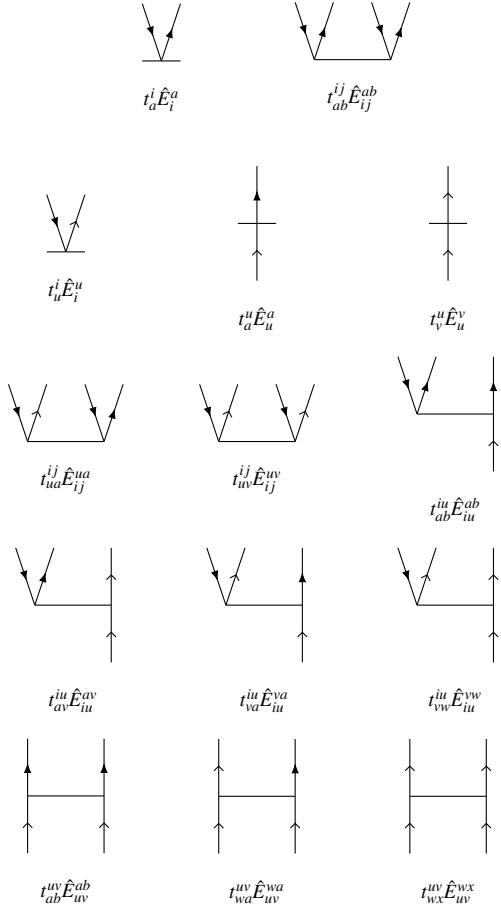
\begin{figure}[hbt!]
    \centering
    \resizebox{!}{0.5\textheight}{\begin{tikzpicture}
\node at (-5,10) {\begin{tikzpicture}[baseline={([yshift=-.5ex]current bounding box.center)}]
\begin{scope}[decoration={
    markings,
    mark=at position 0.58 with {\arrow{Latex[scale=1.5]}}}]
    \draw[ultra thick] (-1,0)--(1,0);
    \draw[ultra thick, postaction={decorate}] (0,0)--(1,3);
    \draw[ultra thick, postaction={decorate}] (-1,3)--(0,0);
\end{scope}
\node at (0,-2) {\fontsize{30}{60}$t_{a}^{i}\hat{E}_{i}^{a}$};
\end{tikzpicture}};
\node at (5,10) {\begin{tikzpicture}[baseline={([yshift=-.5ex]current bounding box.center)}]
\begin{scope}[decoration={
    markings,
    mark=at position 0.58 with {\arrow{Latex[scale=1.5]}}}]
    \draw[ultra thick] (-2,0)--(2,0);
    \draw[ultra thick, postaction={decorate}] (-2,0)--(-1,3);
    \draw[ultra thick, postaction={decorate}] (-3,3)--(-2,0);
    \draw[ultra thick, postaction={decorate}] (2,0)--(3,3);
    \draw[ultra thick, postaction={decorate}] (1,3)--(2,0);
\end{scope}
\node at (0,-2) {\fontsize{30}{60}$t_{ab}^{ij}\hat{E}_{ij}^{ab}$};
\end{tikzpicture}};
\node at (-10,0) {\begin{tikzpicture}[baseline={([yshift=-.5ex]current bounding box.center)}]
\begin{scope}[decoration={
    markings,
    mark=at position 0.58 with {\arrow{Latex[scale=1.5]}}}]
    \draw[ultra thick] (-1,0)--(1,0);
    \draw[ultra thick, postaction={decorate}] (-1,3)--(0,0);
\end{scope}
\begin{scope}[decoration={
    markings,
    mark=at position 0.54 with {\arrow{Straight Barb[scale=1.5]}}}]
    \draw[ultra thick, postaction={decorate}] (0,0)--(1,3);
\end{scope}
\node at (0,-2) {\fontsize{30}{60}$t_{u}^{i}\hat{E}_{i}^{u}$};
\end{tikzpicture}};
\node at (0,0) {\begin{tikzpicture}[baseline={([yshift=-.5ex]current bounding box.center)}]
\begin{scope}[decoration={
    markings,
    mark=at position 0.58 with {\arrow{Latex[scale=1.5]}}}]
    \draw[ultra thick] (-1,0)--(1,0);
    \draw[ultra thick, postaction={decorate}] (0,0)--(0,3);
\end{scope}
\begin{scope}[decoration={
    markings,
    mark=at position 0.54 with {\arrow{Straight Barb[scale=1.5]}}}]
    \draw[ultra thick, postaction={decorate}] (0,-3)--(0,0);
\end{scope}
\node at (0,-5) {\fontsize{30}{60}$t_{a}^{u}\hat{E}_{u}^{a}$};
\end{tikzpicture}};
\node at (10,0) {\begin{tikzpicture}[baseline={([yshift=-.5ex]current bounding box.center)}]
\begin{scope}[decoration={
    markings,
    mark=at position 0.58 with {\arrow{Latex[scale=1.5]}}}]
    \draw[ultra thick] (-1,0)--(1,0);
\end{scope}
\begin{scope}[decoration={
    markings,
    mark=at position 0.54 with {\arrow{Straight Barb[scale=1.5]}}}]
    \draw[ultra thick, postaction={decorate}] (0,0)--(0,3);
    \draw[ultra thick, postaction={decorate}] (0,-3)--(0,0);
\end{scope}
\node at (0,-5) {\fontsize{30}{60}$t_{v}^{u}\hat{E}_{u}^{v}$};
\end{tikzpicture}};
\node at (-10,-10) {\begin{tikzpicture}[baseline={([yshift=-.5ex]current bounding box.center)}]
\begin{scope}[decoration={
    markings,
    mark=at position 0.58 with {\arrow{Latex[scale=1.5]}}}]
    \draw[ultra thick] (-2,0)--(2,0);
    \draw[ultra thick, postaction={decorate}] (-3,3)--(-2,0);
    \draw[ultra thick, postaction={decorate}] (2,0)--(3,3);
    \draw[ultra thick, postaction={decorate}] (1,3)--(2,0);
\end{scope}
\begin{scope}[decoration={
    markings,
    mark=at position 0.54 with {\arrow{Straight Barb[scale=1.5]}}}]
    \draw[ultra thick, postaction={decorate}] (-2,0)--(-1,3);
\end{scope}
\node at (0,-2) {\fontsize{30}{60}$t_{ua}^{ij}\hat{E}_{ij}^{ua}$};
\end{tikzpicture}};
\node at (0,-10) {\begin{tikzpicture}[baseline={([yshift=-.5ex]current bounding box.center)}]
\begin{scope}[decoration={
    markings,
    mark=at position 0.58 with {\arrow{Latex[scale=1.5]}}}]
    \draw[ultra thick] (-2,0)--(2,0);
    \draw[ultra thick, postaction={decorate}] (-3,3)--(-2,0);
    \draw[ultra thick, postaction={decorate}] (1,3)--(2,0);
\end{scope}
\begin{scope}[decoration={
    markings,
    mark=at position 0.54 with {\arrow{Straight Barb[scale=1.5]}}}]
    \draw[ultra thick, postaction={decorate}] (-2,0)--(-1,3);
    \draw[ultra thick, postaction={decorate}] (2,0)--(3,3);
\end{scope}
\node at (0,-2) {\fontsize{30}{60}$t_{uv}^{ij}\hat{E}_{ij}^{uv}$};
\end{tikzpicture}};
\node at (10,-10) {\begin{tikzpicture}[baseline={([yshift=-.5ex]current bounding box.center)}]
\begin{scope}[decoration={
    markings,
    mark=at position 0.58 with {\arrow{Latex[scale=1.5]}}}]
    \draw[ultra thick] (-2,0)--(2,0);
    \draw[ultra thick, postaction={decorate}] (-2,0)--(-1,3);
    \draw[ultra thick, postaction={decorate}] (-3,3)--(-2,0);
    \draw[ultra thick, postaction={decorate}] (2,0)--(2,3);
\end{scope}
\begin{scope}[decoration={
    markings,
    mark=at position 0.54 with {\arrow{Straight Barb[scale=1.5]}}}]
    \draw[ultra thick, postaction={decorate}] (2,-3)--(2,0);
\end{scope}
\node at (0,-5) {\fontsize{30}{60}$t_{ab}^{iu}\hat{E}_{iu}^{ab}$};
\end{tikzpicture}};
\node at (-10,-20) {\begin{tikzpicture}[baseline={([yshift=-.5ex]current bounding box.center)}]
\begin{scope}[decoration={
    markings,
    mark=at position 0.58 with {\arrow{Latex[scale=1.5]}}}]
    \draw[ultra thick] (-2,0)--(2,0);
    \draw[ultra thick, postaction={decorate}] (-3,3)--(-2,0);
    \draw[ultra thick, postaction={decorate}] (-2,0)--(-1,3);
\end{scope}
\begin{scope}[decoration={
    markings,
    mark=at position 0.54 with {\arrow{Straight Barb[scale=1.5]}}}]
    \draw[ultra thick, postaction={decorate}] (2,-3)--(2,0);
    \draw[ultra thick, postaction={decorate}] (2,0)--(2,3);
\end{scope}
\node at (0,-5) {\fontsize{30}{60}$t_{av}^{iu}\hat{E}_{iu}^{av}$};
\end{tikzpicture}};
\node at (0,-20) {\begin{tikzpicture}[baseline={([yshift=-.5ex]current bounding box.center)}]
\begin{scope}[decoration={
    markings,
    mark=at position 0.58 with {\arrow{Latex[scale=1.5]}}}]
    \draw[ultra thick] (-2,0)--(2,0);
    \draw[ultra thick, postaction={decorate}] (-3,3)--(-2,0);
    \draw[ultra thick, postaction={decorate}] (2,0)--(2,3);
\end{scope}
\begin{scope}[decoration={
    markings,
    mark=at position 0.54 with {\arrow{Straight Barb[scale=1.5]}}}]
    \draw[ultra thick, postaction={decorate}] (2,-3)--(2,0);
    \draw[ultra thick, postaction={decorate}] (-2,0)--(-1,3);
\end{scope}
\node at (0,-5) {\fontsize{30}{60}$t_{va}^{iu}\hat{E}_{iu}^{va}$};
\end{tikzpicture}};
\node at (10,-20) {\begin{tikzpicture}[baseline={([yshift=-.5ex]current bounding box.center)}]
\begin{scope}[decoration={
    markings,
    mark=at position 0.58 with {\arrow{Latex[scale=1.5]}}}]
    \draw[ultra thick] (-2,0)--(2,0);
    \draw[ultra thick, postaction={decorate}] (-3,3)--(-2,0);
\end{scope}
\begin{scope}[decoration={
    markings,
    mark=at position 0.54 with {\arrow{Straight Barb[scale=1.5]}}}]
    \draw[ultra thick, postaction={decorate}] (2,-3)--(2,0);
    \draw[ultra thick, postaction={decorate}] (-2,0)--(-1,3);
    \draw[ultra thick, postaction={decorate}] (2,0)--(2,3);
\end{scope}
\node at (0,-5) {\fontsize{30}{60}$t_{vw}^{iu}\hat{E}_{iu}^{vw}$};
\end{tikzpicture}};
\node at (-10,-30) {\begin{tikzpicture}[baseline={([yshift=-.5ex]current bounding box.center)}]
\begin{scope}[decoration={
    markings,
    mark=at position 0.58 with {\arrow{Latex[scale=1.5]}}}]
    \draw[ultra thick] (-2,0)--(2,0);
    \draw[ultra thick, postaction={decorate}] (-2,0)--(-2,3);
    \draw[ultra thick, postaction={decorate}] (2,0)--(2,3);
\end{scope}
\begin{scope}[decoration={
    markings,
    mark=at position 0.54 with {\arrow{Straight Barb[scale=1.5]}}}]
    \draw[ultra thick, postaction={decorate}] (-2,-3)--(-2,0);
    \draw[ultra thick, postaction={decorate}] (2,-3)--(2,0);
\end{scope}
\node at (0,-5) {\fontsize{30}{60}$t_{ab}^{uv}\hat{E}_{uv}^{ab}$};
\end{tikzpicture}};
\node at (0,-30) {\begin{tikzpicture}[baseline={([yshift=-.5ex]current bounding box.center)}]
\begin{scope}[decoration={
    markings,
    mark=at position 0.58 with {\arrow{Latex[scale=1.5]}}}]
    \draw[ultra thick] (-2,0)--(2,0);
    \draw[ultra thick, postaction={decorate}] (2,0)--(2,3);
\end{scope}
\begin{scope}[decoration={
    markings,
    mark=at position 0.54 with {\arrow{Straight Barb[scale=1.5]}}}]
    \draw[ultra thick, postaction={decorate}] (-2,0)--(-2,3);
    \draw[ultra thick, postaction={decorate}] (-2,-3)--(-2,0);
    \draw[ultra thick, postaction={decorate}] (2,-3)--(2,0);
\end{scope}
\node at (0,-5) {\fontsize{30}{60}$t_{wa}^{uv}\hat{E}_{uv}^{wa}$};
\end{tikzpicture}};
\node at (10,-30) {\begin{tikzpicture}[baseline={([yshift=-.5ex]current bounding box.center)}]
\begin{scope}[decoration={
    markings,
    mark=at position 0.58 with {\arrow{Latex[scale=1.5]}}}]
    \draw[ultra thick] (-2,0)--(2,0);
\end{scope}
\begin{scope}[decoration={
    markings,
    mark=at position 0.54 with {\arrow{Straight Barb[scale=1.5]}}}]
    \draw[ultra thick, postaction={decorate}] (-2,0)--(-2,3);
    \draw[ultra thick, postaction={decorate}] (2,0)--(2,3);
    \draw[ultra thick, postaction={decorate}] (-2,-3)--(-2,0);
    \draw[ultra thick, postaction={decorate}] (2,-3)--(2,0);
\end{scope}
\node at (0,-5) {\fontsize{30}{60}$t_{wx}^{uv}\hat{E}_{uv}^{wx}$};
\end{tikzpicture}};
\end{tikzpicture}}
    \caption{Goldstone diagrams for the singles and doubles amplitudes used in our NOECCSD. Core and virtual electrons are represented by solid arrows; active electrons are represented by skeleton arrows.}%{\color{red}Alex - need updated figures with active cases with skeleton arrows }}
    \label{fig:amplitudeDiagrams}
\end{figure}

We choose the cluster operator so that it generates the entire first-order interacting space out of the reference state.
For excitations from core orbitals to virtual, this leads to the usual inclusion of singles and doubles as in standard CCSD approaches.
We also include core-to-active, active-to-virtual, and active-to-active excitations in both the singles and doubles. 
The active-to-active excitations are important to allow the possibility of correlation-induced orbital relaxation of the reference state. The Goldstone diagrams for the general excitation operators for NOECCSD are depicted in Figure \ref{fig:amplitudeDiagrams}.
For reference states with only one active orbital, we exclude purely active-to-active excitations in the singles and doubles.

The active-to-active amplitudes introduce spectator excitations, where a double excitation involves an electron being annihilated and created in the same active orbital and acts as a single excitation when applied to the reference.
The same residual equation is therefore obtained by projection with either the single excitation $\hat{E}_p^q$ or the double excitation with a spectator $\hat{E}_{pu}^{qu}$, and there are an insufficient number of equations to uniquely determine the amplitudes\cite{mukherjee_correlation_1975, jeziorski_coupled-cluster_1981, haque_application_1984, pal_development_1984, lindgren_connectivity_1987}. In the linearised approximation l-NOECCSD, the redundant excitations are interchangeable and the infinite set of amplitudes that solve the residual equations all result in exactly the same energy.  We have also observed that different amplitude solutions also yield exactly the same energies for the q-NOECCSD method, although we do not expect this to be true in general. 
Despite the fact that in the current formulation the equations contain redundant excitations and are under-determined,
%{\color{blue} Nick: For l-NOECC, they have exactly the same energy. As far as I know, we don't yet have numbers for q-NOECC. IF we want to give numbers, it may be more appropriate to do it in the discussion section. I think it's good to mention the infinite family of solutions here, and that DIIS converges to one of them }
we find that is always possible to converge the residual equations. Typically, convergence requires use of an appropriate level shift and the Direct Inversion in the Iterative Subspace (DIIS) technique\cite{pulay_convergence_1980}.
%However we acknowledge that the fully developed theory should employ sufficiency conditions or another procedure to obtain unique amplitudes.
It should be noted that, while spectator excitations are present, we choose to include these only at the ranks denoted by the ansatz, i.e. the NOECCSD cluster operator only includes one- and two-body operators, even though there may be higher rank operators that produce the same configuration from the reference.

We make no attempt at spin completeness of the excitation manifold. For single-reference cases the dominant spin-coupling is present in the CSF and the remaining dynamic correlation processes are contained within the first-order interacting space. The higher-rank excitations necessary for spin completeness lead to computationally expensive terms in the NOECC equations and are not necessary to guarantee spin adaptation. We expect that higher-order excitations will be of much less significance that the ones generating the first-order interacting space.
Hermann and Hanrath found that although excluding spectators in spin-adapted CC treatments gave a significant spin-incompleteness error in the energy, including them only up to the tensor rank was sufficient to remove it almost entirely, despite the excitation manifold not being fully spin-complete.\cite{hermann_analysis_2021}

\section{Method}

\subsection{Automated equation generator}
The large numbers of terms appearing in the spin-free open-shell coupled cluster equations make it necessary to automate the generation of working equations. We therefore constructed an object-oriented Python code that implements a second-quantised operator algebra along with a tensor algebra to represent the associated coefficients. In this framework, a particular coupled-cluster \textit{ansatz} may be represented as a sum of tensor products, with associated second-quantised operator products. The open-shell  reference is included through the application of second-quantised operators with specific orbital indices that create the appropriate configuration state function out of the closed shell Fermi vacuum.
Rigorous spin adaptation of the theory is performed at the level of the coupled-cluster ansatz by including all terms in the spin summation.
This brute force approach leads to very many terms in the equations, but produces a fully spin-adapted coupled cluster theory.
The equations are projected onto the excited state manifold generated by the cluster operators and Wick's theorem is applied to obtain the terms that contribute to the energy and amplitude equations.

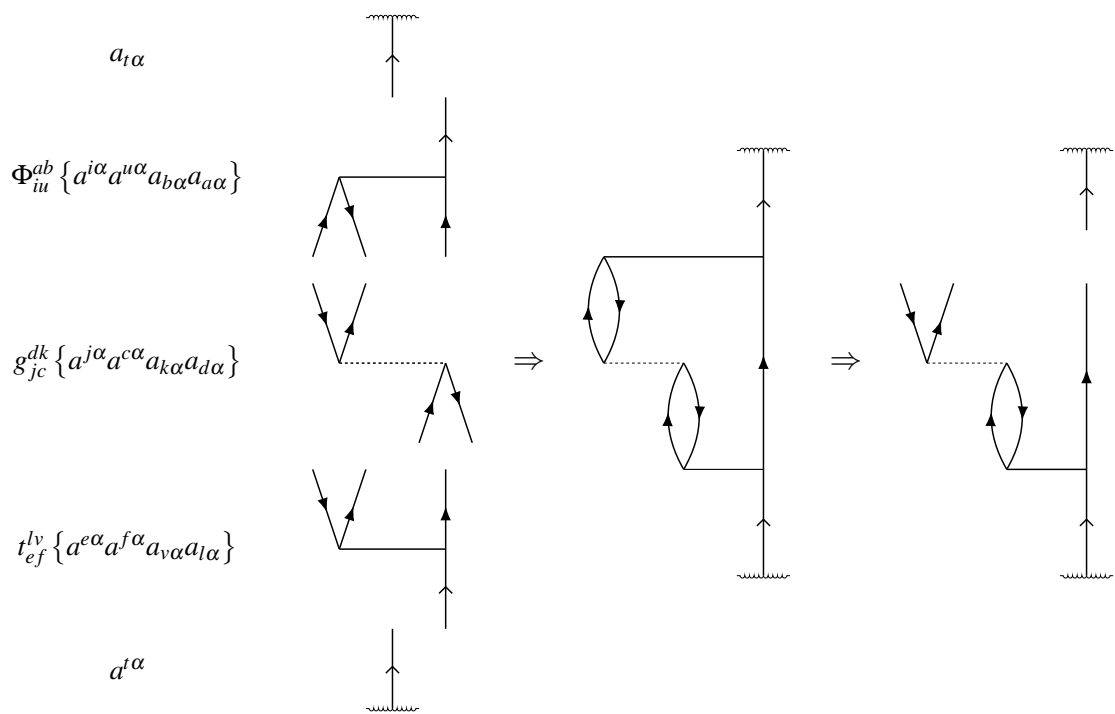
\begin{figure*}[tb!]
    \centering
    \resizebox{0.8\textwidth}{!}{
    $\vcenter{\hbox{\scalebox{0.5}{\begin{tikzpicture}
\begin{scope}[decoration={
    markings,
    mark=at position 0.54 with {\arrow{Latex[scale=1.5]}}}]
    \draw[ultra thick] (-2,0)--(2,0);
    \draw[ultra thick, postaction={decorate}] (-2,0)--(-1,3);
    \draw[ultra thick, postaction={decorate}] (-3,3)--(-2,0);
    \draw[ultra thick, postaction={decorate}] (2,0)--(2,3);
\end{scope}
\begin{scope}[decoration={
    markings,
    mark=at position 0.54 with {\arrow{Straight Barb[scale=1.5]}}}]
    \draw[ultra thick, postaction={decorate}] (2,-3)--(2,0);
\end{scope}
\begin{scope}[decoration={
    markings,
    mark=at position 0.54 with {\arrow{Latex[scale=1.5]}}}]
    \draw[ultra thick] (-2,14)--(2,14);
    \draw[ultra thick, postaction={decorate}] (-2,14)--(-1,11);
    \draw[ultra thick, postaction={decorate}] (-3,11)--(-2,14);
    \draw[ultra thick, postaction={decorate}] (2,11)--(2,14);
\end{scope}
\begin{scope}[decoration={
    markings,
    mark=at position 0.54 with {\arrow{Straight Barb[scale=1.5]}}}]
    \draw[ultra thick, postaction={decorate}] (2,14)--(2,17);
\end{scope}
\begin{scope}[decoration={
    markings,
    mark=at position 0.54 with {\arrow{Latex[scale=1.5]}}}]
    \draw[ultra thick, dashed] (-2,7)--(2,7);
    \draw[ultra thick, postaction={decorate}] (-2,7)--(-1,10);
    \draw[ultra thick, postaction={decorate}] (-3,10)--(-2,7);
    \draw[ultra thick, postaction={decorate}] (2,7)--(3,4);
    \draw[ultra thick, postaction={decorate}] (1,4)--(2,7);
\end{scope}
\begin{scope}[decoration={
    markings,
    mark=at position 0.54 with {\arrow{Straight Barb[scale=1.5]}}}]
    \draw[ultra thick, postaction={decorate}] (0,-6)--(0,-3);
    \draw[ultra thick, postaction={decorate}] (0,17)--(0,20);
\end{scope}
\draw[snake=bumps] (0,20) -- (1,20);
\draw[snake=bumps, mirror snake] (0,20) -- (-1,20);
\draw[snake=bumps] (0,-6) -- (-1,-6);
\draw[snake=bumps, mirror snake] (0,-6) -- (1,-6);
\node[] at (-10, -4.5) {\fontsize{30}{60} $a^{t\alpha}$};
\node[] at (-10, 0) {\fontsize{30}{60} $t_{ef}^{lv}\left\{a^{e\alpha}a^{f\alpha}a_{v\alpha}a_{l\alpha}\right\}$};
\node[] at (-10, 7) {\fontsize{30}{60} $g_{jc}^{dk}\left\{a^{j\alpha}a^{c\alpha}a_{k\alpha}a_{d\alpha}\right\}$};
\node[] at (-10, 14) {\fontsize{30}{60} $\Phi_{iu}^{ab}\left\{a^{i\alpha}a^{u\alpha}a_{b\alpha}a_{a\alpha}\right\}$};
\node[] at (-10, 18.5) {\fontsize{30}{60} $a_{t\alpha}$};
\end{tikzpicture}}}}
    \hspace{20pt}
    \vcenter{\hbox{\scalebox{1}{\Huge$\Rightarrow$}}}
    \hspace{20pt}
    \vcenter{\hbox{\scalebox{0.5}{\begin{tikzpicture}
\begin{scope}[decoration={
    markings,
    mark=at position 0.54 with {\arrow{Latex[scale=1.5]}}}]
    \draw[very thick] (-3,0)--(0,0);
    \draw[dashed] (-6,4)--(-3,4);
    \draw[ultra thick] (-6,8)--(0,8);
    \draw[ultra thick, postaction={decorate}] (-3,0) to [out=120, in=-120] (-3,4);
    \draw[ultra thick, postaction={decorate}] (-3,4) to [out=-60, in=60] (-3,0);
    \draw[ultra thick, postaction={decorate}] (-6,4) to [out=120, in=-120] (-6,8);
    \draw[ultra thick, postaction={decorate}] (-6,8) to [out=-60, in=60] (-6,4);
    \draw[ultra thick, postaction={decorate}] (0,0)--(0,8);
\end{scope}
\begin{scope}[decoration={
    markings,
    mark=at position 0.54 with {\arrow{Straight Barb[scale=1.5]}}}]
    \draw[ultra thick, postaction={decorate}] (0,-4)--(0,0);
    \draw[ultra thick, postaction={decorate}] (0,8)--(0,12);
\end{scope}

\draw[snake=bumps] (0,12) -- (1,12);
\draw[snake=bumps, mirror snake] (0,12) -- (-1,12);
\draw[snake=bumps] (0,-4) -- (-1,-4);
\draw[snake=bumps, mirror snake] (0,-4) -- (1,-4);
\end{tikzpicture}}}}
    \hspace{20pt}
    \vcenter{\hbox{\scalebox{1}{\Huge$\Rightarrow$}}}
    \hspace{20pt}
    \vcenter{\hbox{\scalebox{0.5}{\begin{tikzpicture}
\begin{scope}[decoration={
    markings,
    mark=at position 0.54 with {\arrow{Latex[scale=1.5]}}}]
    \draw[ultra thick] (-3,0)--(0,0);
    \draw[dashed] (-6,4)--(-3,4);
    \draw[ultra thick, postaction={decorate}] (-3,0) to [out=120, in=-120] (-3,4);
    \draw[ultra thick, postaction={decorate}] (-3,4) to [out=-60, in=60] (-3,0);
    \draw[ultra thick, postaction={decorate}] (-6,4)--(-5,7);
    \draw[ultra thick, postaction={decorate}] (-7,7)--(-6,4);
    \draw[ultra thick, postaction={decorate}] (0,0)--(0,7);
\end{scope}
\begin{scope}[decoration={
    markings,
    mark=at position 0.54 with {\arrow{Straight Barb[scale=1.5]}}}]
    \draw[ultra thick, postaction={decorate}] (0,-4)--(0,0);
    \draw[ultra thick, postaction={decorate}] (0,9)--(0,12);
\end{scope}
\draw[snake=bumps] (0,12) -- (1,12);
\draw[snake=bumps, mirror snake] (0,12) -- (-1,12);
\draw[snake=bumps] (0,-4) -- (-1,-4);
\draw[snake=bumps, mirror snake] (0,-4) -- (1,-4);
\end{tikzpicture}}}}$
    }
    \caption{Goldstone diagram construction representing the automated generation of the residual term $r^{iu}_{ab}=g_{jc}^{dk}\,t_{ef}^{lv}\,\delta^{u}_{t}\delta^{i}_{k}\delta^{f}_{b}\delta^{c}_{a}\delta^{j}_{l}\delta^{e}_{d}\delta^{t}_{v}$}
    \label{fig:equationGeneratorExample}
\end{figure*}

As an example, consider a doublet with one open-shell electron, with spin $\alpha$, in an orbital labelled $t$.
Each contribution $r_{I}$ to a residual $R_{I}$ is found by evaluating its contribution to the expectation value when projected onto the excitation manifold, 
\begin{equation}
    \Braket{a_{t\alpha}|\hat{\Phi}_{I} \hat E_I^\dagger r_{I}|a^{t\alpha}}
\end{equation}
For example, the linear contribution $r^{iu}_{ab}$ of amplitude $t^{iu}_{ab}$ with  $g^{bj}_{ia}$ to the residual $R^{iu}_{ab}$ is found from

\begin{equation}
    \Braket{a_{t\alpha}|\phi_{iu}^{ab}\,\left\{\hat E^{iu}_{ba}\right\}g_{jc}^{dk}\,\left\{\hat E^{jc}_{kd}\right\}t_{ef}^{lv}\,\left\{\hat E^{ef}_{vl}\right\}|a^{t\alpha}}
\end{equation}
One of the spin cases is
{\small
\begin{equation}
    \Braket{a_{t\alpha}|\phi_{iu}^{ab}\,\left\{a^{i\alpha}a^{u\alpha}a_{b\alpha}a_{a\alpha}\right\}g_{jc}^{dk}\,\left\{a^{j\alpha}a^{c\alpha}a_{k\alpha}a_{d\alpha}\right\}t_{ef}^{lv}\,\left\{a^{e\alpha}a^{f\alpha}a_{v\alpha}a_{l\alpha}\right\}|a^{t\alpha}}
\end{equation}
}
where possible full contraction is
\begin{equation}
    \phi_{iu}^{ab}\,r^{iu}_{ab}=    \phi_{iu}^{ab}\,g_{jc}^{dk}\,t_{ef}^{lv}\,\delta^{u}_{t}\delta^{i}_{k}\delta^{f}_{b}\delta^{c}_{a}\delta^{j}_{l}\delta^{e}_{d}\delta^{t}_{v}
\end{equation}
for a residual contribution of 
\begin{equation}
r^{iu}_{ab}=g_{jc}^{dk}\,t_{ef}^{lv}\,\delta^{u}_{t}\delta^{i}_{k}\delta^{f}_{b}\delta^{c}_{a}\delta^{j}_{l}\delta^{e}_{d}\delta^{t}_{v}
\end{equation}
The automated procedure to generate this expression is equivalent to the Goldstone diagram construction in figure \ref{fig:equationGeneratorExample}.
% \begin{figure}[h!]
%     \centering
%     \resizebox{0.9*\textwidth}{!}{
%     $\vcenter{\hbox{\scalebox{0.3}{\input{diagrams/uncontracteddiagram.tex}}}}
%     \hspace{20pt}
%     \vcenter{\hbox{\scalebox{1}{\Huge$\Rightarrow$}}}
%     \hspace{20pt}
%     \vcenter{\hbox{\scalebox{0.5}{\input{diagrams/contracteddiagram.tex}}}}
%     \hspace{20pt}
%     \vcenter{\hbox{\scalebox{1}{\Huge$\Rightarrow$}}}
%     \hspace{20pt}
%     \vcenter{\hbox{\scalebox{0.5}{\input{diagrams/residualdiagram}}}}$}
%     \caption{Goldstone diagram construction representing the automated generation of the residual term $r^{iu}_{ab}=g_{jc}^{dk}\,t_{ef}^{lv}\,\delta^{u}_{t}\delta^{i}_{k}\delta^{f}_{b}\delta^{c}_{a}\delta^{j}_{l}\delta^{e}_{d}\delta^{t}_{v}$}
%     \label{fig:equationGeneratorExample}
% \end{figure}

This and other possible spin components are accounted for automatically, together with the factor of 4 associated with this diagram.
In our pilot implementation, we do not identify equivalent terms in the equations by the topologies of their diagrammatic construction.
Furthermore, the equations are not factorized in this naive implementation, giving a suboptimal scaling of computational cost that remains to be addressed in future versions of the code.
It is expected that the efficiency will be improved by a substantial factor when these steps are introduced through future code developments.

\subsection{Solution of CC Equations}
The solution scheme used for these NOECC equations is based on the iterative quasi-Newton method commonly employed in coupled cluster methods.
With no true spin-free Fock matrix available, the preconditioner was the generalized Fock matrix, given by
\begin{align}
    F^{m}_{n}&=\sum_{q}\braket{\Phi_{0}|\hat{E}^{m}_{q}|\Phi_{0}}h_{n}^{q}+\sum_{qrs}\braket{\Phi_{0}|\hat{E}^{mr}_{qs}|\Phi_{0}}g_{nr}^{qs}
%    &=\sum_{q}{D}^{m}_{q}h_{n}^{q}+\sum_{qrs}{d}^{mr}_{qs}g_{nr}^{qs}
\end{align}
Iterative solution of the generated equations is aided by use of a level shift, chosen to be larger than the difference between the Fock matrix eigenvalues corresponding to the highest core orbital and lowest virtual orbital. Direct Inversion in the Iterative Subspace (DIIS) was used to accelerate convergence\cite{pulay_convergence_1980, scuseria_accelerating_1986}. 

\section{Results}

\subsection{Doublets}
We start with doublet electronic states because the single determinant ROHF reference can be correlated using standard unrestricted and spin-restricted spin-orbital coupled cluster implementations, which enables us to
compare directly to our spin-free approach.
In Table \ref{tab:doubletECorr} we present results for unrestricted CCSD (uCCSD), Szalay and Gauss's spin-restricted CCSD (rCCSD)\cite{szalay_spin-restricted_1997} and our normal-ordered exponential NOECCSD, truncated to linear and quadratic order. Szalay and Gauss's 
spin-restricted approach constrains the cluster amplitudes by imposing the exact $S^2$ expectation value, rather than fully spin-adapting the amplitudes. We also present results for the quadratic spin-free CCSD without normal-ordering. 
The geometries were obtained from Szalay and Gauss\cite{szalay_spin-restricted_1997} and we use the same cc-pVTZ basis as that work. In all cases, the ROHF state is used as the reference determinant. Energies were converged to $10^{-10}$ $\mathrm{E_{H}}$.

\begin{table}[h]
        \centering
        \resizebox{\columnwidth}{!}{
        \begin{tabular}{c|c|c|c|c|c}
            Molecule & uCCSD & rCCSD &  l-NOECCSD &  q-NOECCSD &  q-CCSD \\
            \hline
             $\mathrm{{}^{2}CH}$     &$ -141.105 $&$ -140.973 $&$ -137.013 $&$ -140.851 $&$ -141.049$\\
             $\mathrm{{}^{2}OH}$     &$ -230.302 $&$ -230.156 $&$ -223.133 $&$ -230.306 $&$ -230.331$\\
             $\mathrm{{}^{2}CN}$     &$ -354.724 $&$ -353.995 $&$ -325.369 $&$ -353.749 $&$ -354.454$\\
             $\mathrm{{}^{2}NO}$     &$ -433.834 $&$ -433.354 $&$ -402.668 $&$ -433.327 $&$ -434.265$\\
             $\mathrm{{}^{2}CH_{3}}$ &$ -199.002 $&$ -198.832 $&$ -191.233 $&$ -198.833 $&$ -198.974$\\
             $\mathrm{{}^{2}NH_{2}}$ &$ -220.008 $&$ -219.831 $&$ -211.911 $&$ -219.806 $&$ -219.991$
        \end{tabular}}
        \caption{l-NOECCSD and q-NOECCSD correlation energies in $\mathrm{mE_{H}}$ for ROHF doublets states in a cc-pVTZ basis compared to  unrestricted spin-orbital based CCSD (uCCSD), Szalay and Gauss's restricted CCSD (rCCSD) and spin-free quadratic CCSD (q-CCSD).}
        \label{tab:doubletECorr}
    \end{table}

The rCCSD correlation energies are smaller than uCCSD due to the spin-restriction imposed on the amplitudes. The spin-free NOECCSD energies are in close agreement with rCCSD, differing by only 0.3 m$\mathrm{E_{H}}$. 
Exact agreement between spin-restricted and spin-adapted theories is not expected, even if we had not truncated the NOECCSD equations to quadratic order. The difference between the two approaches is small in comparison to the expected magnitude of the contribution from the three-body excitations that are absent in both methods. 
The non normal-ordered spin-free CCSD correlation energies are universally lower than for NOECCSD, in some cases by as much as 0.8 m$\mathrm{E_{H}}$. 
This is because, without normal-ordering, quadratic terms involving active orbitals result in spurious additional energy contributions. For example $t_a^u t_u^i E_u^a E_i^u$ gives rise to an additional core to virtual excitation $t_a^u t_u^i E_i^a$ that lowers the energy. 
Normal-ordering has no impact on the linear terms, but linearised NOECCSD only recovers 90\% of the correlation energy. 

%Even when including only terms linear in the excitation operator, the spin-free CCSD equations recover over 90\% of the correlation energy obtained by the full spin-orbital calculation.
%When quadratic terms are included, the correlation energy differs from the full spin-orbital calculation by less than 0.1\%.
%The neglect of contractions in the normal-ordered exponential version of this calculation causes a discrepancy that is still less than 0.3\% of the correlation energy.
%Although the spin restricted formulation of Szalay and Gauss generally recovers slightly more of the correlation energy, it does not give a fully spin-adapted wavefunction, on which account all three of these new results are preferable.
% We expect a difference in the energies between the proposed spin-adapted NOECC and the spin-restricted CC method of Szalay and Gauss. In the spin-restricted formalism, the 

%Although the spin restricted formulation of Szalay and Gauss generally recovers slightly more of the correlation energy, NOECC offers the benefit of being fully spin adapted {\color{red} and therefore ...}.
%For these small systems, agreement at the sub-$\mathrm{mE_{H}}$ level between spin-adapted and spin-restricted methods can be achieved even with only terms up to quadratic order in the amplitudes.%, and even without any contractions between them.

%{\color{red} highlight any conclusions we carry forwards to next sections? - e.g. do not present any more non-normal ordered results}

\subsection{Open Shell Singlets and Triplets}

\subsubsection{Beryllium Atom} 

Having demonstrated our spin-free NOECC framework for one-electron doublets, we now turn to the two-electron cases: the triplet and the open-shell singlet.
The ${}^3P$ and ${}^1P$ 2s2p excited states of the beryllium atom exemplify these electronic configurations, in a system simple enough to allow comparison of the singlet-triplet splitting with full configuration interaction (FCI).

Table \ref{tab:BeSplitting} lists energies computed for the CSF references, and all-electron FCI and NOECCSD energies of the singlet ${}^1P$ and triplet ${}^3P$ states. uCCSD energies are also included for the triplet state, which has a single reference determinant.
We use the cc-pCVTZ basis and the orbitals for both CSF references were generated from an ROHF calculation for the $M=1$ triplet state. The CSF for the open-shell singlet is that generated by Eq. \ref{eqn:csf-singlet}. Energies are converged to $10^{-10}$ $\mathrm{E_{H}}$.

    \begin{table}[h]
        \centering
        \resizebox{\linewidth}{!}{
        \begin{tabular}{c|r|r|r|r|r}
            $2S+1$& CSF  & uCCSD &  l-NOECCSD &  q-NOECCSD & FCI  \\
            \hline
            $3$ &$ -14.513129 $&$ -14.561352 $&$ -14.561504 $&$ -14.561358 $&$ -14.561412 $\\
            $1$ &$ -14.356425 $&$ -          $&$ -14.473941 $&$ -14.462727 $&$ -14.463057 $\\
            \hline
            $\Delta$ &$ 0.156703 $&$ - $&$ 0.087563 $&$ 0.098630 $&$ 0.098355 $
        \end{tabular}}
        \caption{l-NOECCSD and q-NOECCSD $2s2p$ excited state energies and singlet-triplet gap $\Delta$ of the beryllium atom in $\mathrm{E_{H}}$ in a cc-pCVTZ basis, compared to the energies of the CSF references, the full CI energies, and the unrestricted CCSD energy for the triplet state.}
        \label{tab:BeSplitting}
    \end{table}

The uCCSD, q-NOECCSD and FCI energies for the single-reference triplet state are all in excellent agreement, within 60 $\mu\mathrm{E_{H}}$. The agreement between CCSD and FCI is a consequence of the higher-order excitations all involving core-valence correlation, which is small for Be.
We can also conclude that spin-contamination in the uCCSD wavefunction for this state is very small, and that the neglected cubic and higher NOECCSD contributions must also either be small in magnitude, or cancel. 

%The agreement of the spin-free NOECCSD energies with the spin-orbital calculation for the triplet improves from 0.15 m$\mathrm{E_{H}}$ to only a few $\mu\mathrm{E_{H}}$ when going from the linear to quadratic truncation.
%While the quadratic spin-free CCSD and NOECCSD energies for the triplet agree with FCI to within 0.06 m$\mathrm{E_{H}}$, agreement in the singlet energies with FCI is only to within 0.4 m$\mathrm{E_{H}}$.
%The singlet-triplet gap is thus obtained to within 0.4 m$\mathrm{E_{H}}$ of FCI.

The singlet CSF reference is almost 0.1$\mathrm{E_{H}}$ above FCI, whereas the triplet reference was 0.05$\mathrm{E_{H}}$ above the corresponding FCI energy. This is in part a consequence of using the ROHF orbitals optimised for the triplet, but also because the correlation among singlet spin-coupled electrons is larger than triplet spin-coupled due to the Fermi heap in their joint probability distribution.\cite{Tew2007} The effect of both orbital relaxation and correlation are
well captured in the spin-free NOECCSD approach. At linear order, l-NOECCSD is within 10 m$\mathrm{E_{H}}$ of FCI, which reduces to 0.3 m$\mathrm{E_{H}}$
for q-NOECCSD. The singlet-triplet gap is also accurate to 0.3 m$\mathrm{E_{H}}$ compared to FCI.

%The dynamic correlation in the multi-determinantal open shell singlet is obtained with no more difficulty than single determinant high-spin triplet.
%Comparing to the full CI values, the singlet-triplet energy splitting in the reference states is significantly too large, as expected with orbitals that were optimised for the triplet state in a ROHF calculation. On application of spin-free coupled cluster methods, the gap decreased sharply. While a linear ansatz underestimates the singlet-triplet gap, inclusion of quadratic terms in either q-NOECCSD or q-CCSD (the former does not include contractions between different cluster operators) results in a gap that differs from the FCI result by less than 0.4\%. The energies of the open-shell singlet and triplet also agrees with FCI at the sub-$\mathrm{mE_{H}}$ level when quadratic terms are included.
%Although the energy gap between reference states is too large due to the choice of orbitals, the inclusion of single excitations leads us to believe that the NOECCSD energy will depend only weakly on the choice of orbitals.

A fully spin-adapted theory should recover the same energy for each of the $(2S+1)$ spin-projections of a state with total spin $S$. Our approach enables us to correlate arbitrary CSFs, and as a test of our method, we computed all-electron l-NOECCD energies of the $M=1$ and $M=0$ components of the triplet state. The $M=1$ CSF is a single ROHF determinant, whereas the $M=0$ CSF is a linear combination of two determinants (Eq. \ref{eqn:csf-triplet}). The two calculations converged with energies that were identical at every iteration, confirming that our method is fully spin-adapted.
%For the simplest case of linear CCD, the automatically generated equations preserve the degeneracy of the triplet between single and multi-determinantal references.
%{\color{red}Alex - any update on this? is it also true for l-NOECCSD or for q-NOECCSD?}

\subsubsection{Oxygen Molecule}

Another archetypical singlet-triplet system is provided by the oxygen molecule.
%^To examine if the accuracy of our spin-free NOECC is maintained in a larger system, we look to the low-lying states of the oxygen molecule. 
The lowest energy singlet state ${}^{1}\Delta_{g}$ lies 0.0359 $\mathrm{E_{H}}$ above the triplet  ${}^{3}\Sigma_{g}^{-}$ ground state. Both states  correspond to the open-shell configuration $\pi_{g,x}^1\pi_{g,y}^{1}$. The $O_2^{1,1}$ reference CSF for the ${}^{3}\Sigma_{g}^{-}$ is a single ROHF determinant, whereas the $O_2^{0,0}$ reference CSF for the ${}^{1}\Delta_{g}$ state is a linear combination of two determinants.
%We further demonstrate the spin-adapted NOECC framework in a multireference case on a larger system with the lowest singlet excited states of the oxygen molecule.
%The splitting between the ${}^{3}\Sigma_{g}^{-}$ ground state and two low-lying singlet states is investigated; one open-shell, the other closed-shell.
Table \ref{tab:O2Splitting} reports all-electron NOECCSD energies for the singlet and triplet states of $\mathrm{O_{2}}$ at a bond length of 1.2075\AA\ using a cc-pCVTZ basis. For both CSFs, the orbitals were obtained from a ROHF calculation on the triplet state. We also report energies for the  ${}^{3}\Sigma_{g}^{-}$ ground state computed by ROHF-uCCSD for comparison. All energies are converged to $10^{-10}$ $\mathrm{E_{H}}$.

%as calculated for the closed and open-shell CSF references, and the correlated energies using NOECCSD, truncated to linear and quadratic order, alongside the uCCSD for the single-determinant reference states.

%The orbitals were obtained from a triplet ROHF calculation with no frozen core in a cc-pCVTZ basis.
%The high-spin triplet reference state for the NOECC calculation kept the same $\pi_{x}^{*}(1)\pi_{y}^{*}(2)$ active orbital configuration.
%One of the two singlet states has the active orbital configuration given by $\frac{1}{\sqrt{2}}\left(\pi_{x}^{*}(1)\pi_{y}^{*}(2)+\pi_{y}^{*}(1)\pi_{x}^{*}(2)\right)$ active orbital configuration, and the other $\left(\pi_{x}^{*}\pi_{x}^{*}\right)$.
%The first of those states, transforming as $2xy$ in $D_{\infty h}$, corresponds to a ${}^{1}\Delta_{g}$ state.
%The second, transforming as $x^{2}$, does not preserve spatial symmetry between the $x$ and $y$ coordinates and in fact forms one part of the multideterminantal ${}^{1}\Sigma_{g}^{+}$ state transforming as $x^{2}+y^{2}$ that will be the subject of future work.

%Using these three reference states, the linear and quadratic NOECC calculations were converged to a total squared amplitude less than $10^{-16}$ and energy change less than $10^{-12}$, with level shift of 2 and DIIS subspace of 12 iterations in all cases.

    \begin{table}[h]
        \centering
        \resizebox{\linewidth}{!}{
        \begin{tabular}{c|c|c|c|c}
            ${}^{2S+1}\Lambda$& Reference Energy & uCCSD & l-NOECCSD & q-NOECCSD\\
            \hline
            ${}^{3}\Sigma_{g}^{-}$ & $-149.653208$ & $-150.223429$ & $-150.238029 $&$ -150.222443 $\\
            ${}^{1}\Delta_{g}$     &$ -149.605647 $&$ -           $&$ -150.204043 $&$ -150.183136 $\\
            ${}^{1}\Sigma_{g}^{+}$ &$ -149.558086 $&$ - $&$ -150.187127 $&$ -150.156285$\\
            \hline
            ${}^{1}\Delta_{g} - {}^{3}\Sigma_{g}^{-}$ &$ 0.047561 $&$ - $&$ 0.033986 $&$ 0.039307$\\
            ${}^{1}\Sigma_{g}^{+}-{}^{3}\Sigma_{g}^{-}$ & $0.095122$&$ - $&$ 0.0509029 $&$0.066158$
        \end{tabular}}
        \caption{l-NOECCSD and q-NOECCSD energies in $\mathrm{E_{H}}$ of the triplet ground state and lowest-lying open and closed shell singlet excited states of $\mathrm{O_{2}}$ in a cc-pCVTZ basis set, compared to the CSF reference eneriges and the unrestricted CCSD energy of the triplet state.}
        \label{tab:O2Splitting}
    \end{table}
As before, the q-NOECCSD energy is within 1 m$\mathrm{E_{H}}$ of the uCCSD energy, where they can be compared. The computed singlet-triplet gap of 0.0393 $\mathrm{E_{H}}$ is in good agreement with the experimental value of 0.0359 $\mathrm{E_{H}}$, considering that the experimental value is the adiabatic 0-0 energy, whereas we have computed the vertical electronic energy, and that triple excitations are not fully accounted for in our method. The ${}^{1}\Sigma_{g}^{+}$ state, which lies
0.0598 $\mathrm{E_{H}}$ above the triplet ground state, is a $\pi_{g,x}^2 + \pi_{g,y}^{2}$ configuration that is a symmetry-adapted linear combination of two closed-shell CSFs. Our single-reference open-shell formalism extends straightforwardly to this multi-determinant state and we obtain a singlet-triplet gap with q-NOECCSD of 0.0662 $\mathrm{E_{H}}$.

%We have also extended our approach to treat the ${}^{1}\Sigma_{g}^{+}$ state, which is a $\pi_{g,x}^2 + \pi_{g,y}^{2}$ configuration lying 0.0598 $\mathrm{E_{H}}$ above the triplet ground state.
%This state also cannot be treated through uCCSD approaches, but we can nonetheless treat this symmetry-defined linear combination of two closed-shell CSFs as a single multi-determinantal reference.
%We generated new working equations for this configuration and solved them as before, computing the second singlet-triplet gap with q-NOECCSD to be 0.0662 $\mathrm{E_{H}}$.
%, is again only a few m$\mathrm{E_{H}}$ larger than the experimental value of 0.0598 $\mathrm{E_{H}}$.

%The spin-free NOECCSD energy of the triplet, when quadratic terms are included, differs from the spin-orbital CCSD by less than one mEH.
%The discrepancy is expected to arise from contractions between the amplitudes, as well as the cubic and higher order terms in the exponential.

%As before, the spin-free CCSD energies for single determinant references agree with the spin-orbital calculations to within $1mE_{H}$ when quadratic terms are included, demonstrating the possibility of achieving chemical accuracy for systems with two electrons in active orbitals.

\section{Discussion}

We discuss the characteristics of the NOECC method using criteria set out by
K\"ohn and coworkers for MRCC theory: 
size extensivity; 
size consistency;
orbital invariance;
compatibility with SRCC;
satisfying the proper residual condition.

\subsection{Size extensivity}

% The NOECC equations are formulated such that, at each order in the amplitudes, the unlinked terms in the residual equations exactly cancel. 
As shown in Appendix \ref{A-Connectedness}, solving for the amplitude equations is equivalent to solving for a corresponding set of equations comprising only connected terms.
%The argument relies on the fact that $\hat{H} \{ e^{\hat{T}} \}$ can be expressed as the product of $\{ e^{\hat{T}} \}$ with a sum of fully connected terms.
Since the energy of the CSF reference is also size extensive, our NOECC method is fully size extensive when all terms are included up to the finite order at which the equations terminate.
Truncating our equations at a lower order introduces a size-extensivity error.

\subsection{Size consistency}

Size consistency requires that the energy of a molecule composed of two infinitely separated spin-coupled fragments is equal to the sum of the energies of the isolated fragments.
For size consistency to be satisfied, the reference CSF must dissociate into
the corresponding CSFs for the constituent fragments, and non-vanishing higher-order excitations present in the NOECC wave operator for the molecule must be representable through products of lower-order excitations of the fragments. 

The appropriate CSF for spin-coupled, but non-interacting fragments is the one constructed from Clebsch--Gordon coupling of the spins of the fragments. For example, a two-electron open-shell triplet and singlet are formed by high-spin or low-spin coupling two one-electron doublet states, respectively. To assess the size consistency of the NOECC ansatz, we examine the energies for the homolytic bond fission of Li$_2$ as the simplest non-trivial test case.

% A fundamental design feature of coupled-cluster methods is the separability of the wavefunction when treating non-interacting subsystems, and the consequent size-consistency of the correlation energy.
% For multiple copies of a closed shell system, NOECC trivially maintains size consistency by its equivalence to the standard spin-orbital based coupled cluster. For an open-shell system, size consistency demands that the energy of many non-interacting open-shell systems is the same as the sum of the energies of each system, irrespective of the relative spin-coupling between the systems.
% We have not yet been able to prove size consistency in the general case, but in the following
%In the NOECC framework, it must be shown that the use of different sets of working equations for the different subsystems and combined system does not seriously break size consistency.

%\subsubsection{Dissociated Lithium Dimer}

At dissociation, the $^1\Sigma_g^+$ and $^3\Sigma_u^+$ states of lithium dimer should have the same energy, and this should be equal to twice that of an isolated lithium atom in the $^2S$ state. The results of all-electron spin-free NOECCD and NOECCSD calculations assessing the extent to which this is true are presented in Tables \ref{tab:Li2CCD} and \ref{tab:Li2CCSD}, respectively. The orbitals of the $^3\Sigma_u^+$ state of Li$_2$ and the $^2S$ state of the Li atom were obtained from ROHF calculations. The $^1\Sigma_g^+$ state was constructed using the localised $1s$ and $2s$ orbitals of the $^3\Sigma_u^+$ state, ensuring that the energy expectation value of the open-shell singlet CSF exactly matches the ROHF energy of the triplet, and that both are exactly twice that of the doublet.
The cc-pCVTZ basis was used and an interatomic distance of $10^9$\AA\ was chosen to represent the system of the two dissociated, but still spin-coupled, lithium atoms. All energies are converged to $10^{-10}$ $\mathrm{E_{H}}$. uCCD and uCCSD energies are also provided for comparison.
%{\color{green}AG 29/2: linear CCD results exactly replicated (to $10^{-12}$ $\mathrm{E_{H}}$) for both singlet and triplet in the duplicated atomic orbital basis; qNOECC to come}
%{\color{red} do you mean the numbers in the table remain as they are?}

%A $\mathrm{Li}_{2}$ molecule with an interatomic distance of $10^9\AA$ was used to represent the system of two isolated, but still spin-coupled, lithium atoms.
%For the individual lithium atom, the reference was generated by a ROHF calculation with no frozen core in a cc-pCVTZ basis.
%For the dimer a triplet ROHF calculation was performed.
% The triplet reference used the $\mathrm{2s^{\alpha}2p_{z}^{\alpha}}$ configuration and the singlet used $\left(2s^{\alpha}2p_{z}^{\alpha} - 2s^{\alpha}2p_{z}^{\alpha}\right)/\sqrt{2}$ in the same orbitals.
%The high-spin triplet and open-shell singlet, constructed with the $2s$ and $2p$ orbitals found from the ROHF calculation, were used as the reference states.
%The spin-free NOECCD and NOECCSD calculations were converged to a total squared amplitude less than $10^{-16}$ and energy change less than $10^{-12}$, with level shift of 2 and DIIS subspace of 12 iterations.

\begin{table}[h]
        \centering
        \resizebox{0.9\linewidth}{!}{
        \begin{tabular}{c|r|r|r}
              & uCCD & l-NOECCD & q-NOECCD\\
             \hline
             $\mathrm{Li} \, ({}^2S)$                &$ -41.505947 $&$ -41.680942 $&$ -41.532939$ \\
             $\mathrm{Li_{2}}\, ({}^3\Sigma_g^-), r=10^9$\AA &$ -83.011893 $&$ -83.389282 $&$ -83.091473$ \\
             $\mathrm{Li_{2}}\, ({}^1\Sigma_g^+), r=10^9$\AA &$ -          $&$ -83.389289 $&$ -83.091788$ \\
             \hline
             $2E\left({}^2S\right)-E\left({}^3\Sigma_g^-\right)$ &$ 0.000000 $&$ 0.027398 $&$ 0.025595$ \\
             $2E\left({}^2S\right)-E\left({}^1\Sigma_g^+\right)$ &$ - $&$ 0.027404 $&$ 0.025910$
        \end{tabular}}
        \caption{l-NOECCD and q-NOECCD correlation energies in $\mathrm{mE_{H}}$ for $\mathrm{Li}$ and $\mathrm{Li_{2}}$ at a separation of $10^{9}$\AA\ in cc-pCVTZ basis, compared to unrestricted CCD.}
        \label{tab:Li2CCD}
    \end{table}

    \begin{table}[h]
        \centering
        \resizebox{0.9\linewidth}{!}{
        \begin{tabular}{c|r|r|r}
              & uCCSD &  l-NOECCSD &  q-NOECCSD\\
             \hline
             $\mathrm{Li} \, ({}^2S)$                &$ -41.546366 $&$ -41.694641 $&$ -41.545784 $\\
             $\mathrm{Li_{2}}\, ({}^3\Sigma_g^-), r=10^9$\AA &$ -83.092731 $&$ -83.389282 $&$ -83.091498 $\\
             $\mathrm{Li_{2}}\, ({}^1\Sigma_g^+), r=10^9$\AA & -          &$ -83.389289 $&$ -83.091947 $\\
             \hline
             $2E\left({}^2S\right)-E\left({}^3\Sigma_g^-\right)$ &$ 0.000000 $&$ 0.000001 $&$ -0.000070$ \\
             $2E\left({}^2S\right)-E\left({}^1\Sigma_g^+\right)$ & - &$ 0.000007 $&$ 0.000379$
        \end{tabular}}
        \caption{l-NOECCSD and 1-NOECCSD correlation energies in $\mathrm{mE_{H}}$ for $\mathrm{Li}$ and $\mathrm{Li_{2}}$ at a separation of $10^{9}$\AA\ in cc-pCVTZ basis, compared to unrestricted CCSD.}
        \label{tab:Li2CCSD}
    \end{table}

%In both CCD and CCSD, the correlation energies of the lithium atom in its doublet ground state as calculated using the linear and normal-ordered quadratic \textit{ans\"atze} are reasonable, being within $0.2\mathrm{mE_{H}}$ of the equivalent spin-contaminated values.
%For the spin-coupled dimer at separation of $10^9\AA$, reasonable correlation energies were also obtained, within $0.4\mathrm{mE_{H}}$ of the spin-contaminated values.

% {\color{red} for the triplet state uCCD and UCCSD are size consistent as expected.}

% {\color{red} l-NOECCD and q-NOECCD are not size consistent. This needs to be explained.}

% {\color{red} l-NOCCSD and q-NOECCSD are not strictly size consistent, but the inconsistency error is extremely small. - again, this needs to be explained. Do we expect it to be exactly zero if we go beyond quadratic?}

%{\color{green} DPT a reviewer might ask what the symmetry-broken uCCD and uCCSD method gives - this is where a spin-orbital code is used, but with one-half of the singlet CSF state as a single determinant reference - presumably this is also size consistent? This would be a bit of a red herring though, because that approach only works for non-interacting open-shell orbitals.}

% {\color{red} Text below contains interesting information, but the it is not currently linked to the question: is the method size consistent - I have left it here for now}

Both l-NOECCD and q-NOECCD show significant size-inconsistency errors of $0.03$ $\mathrm{mE_{H}}$. Size consistency is violated because the cluster operator includes terms that excite fragment A with spectator orbitals on fragment B, and vice versa
\begin{equation}
    \hat{E}_{pu}^{qu}   \rightarrow \hat{E}_{p_{A}u_{A}}^{q_{A}u_{A}} + \hat{E}_{p_{A}u_{B}}^{q_{A}u_{B}} + \hat{E}_{p_{B}u_{A}}^{q_{B}u_{A}} + \hat{E}_{p_{B}u_{B}}^{q_{B}u_{B}}
\end{equation}
The terms $\hat{E}_{p_{A}u_{B}}^{q_{A}u_{B}} + \hat{E}_{p_{B}u_{A}}^{q_{B}u_{A}}$ are non-vanishing in the molecule, but are absent in the cluster operators of fragments and the cluster operator is therefore not additively separable in the dissociative limit, which for CCD would require
\begin{equation}
    \hat{E}_{pu}^{qu}   \rightarrow \hat{E}_{p_{A}u_{A}}^{q_{A}u_{A}} + \hat{E}_{p_{B}u_{B}}^{q_{B}u_{B}}
\end{equation}
The doubles spectator excitations correspond to single excitations when acting on the reference. Since the l-NOECCSD and q-NOECCSD methods explicitly include the single excitations
$\hat{E}_{p_{A}}^{q_{A}}$, $\hat{E}_{p_{B}}^{q_{B}}$ that were absent in NOECCD, size consistency can be restored numerically through the combined action of the singles and doubles. The size-inconsistency errors are reduced to 7 $\mathrm{nE_H}$ for l-NOECCSD and 4 $\mu\mathrm{E_H}$ for q-NOECCSD. While the NOECCSD method is not rigorously size consistent, it is numerically very close to being size consistent in practice. Similar observations have been made by Hanauer and K\"ohn\cite{hanauer_pilot_2011} for the ic-MRCC method, where they found that the magnitude of size-inconsistency errors depended critically on the way they removed the redundancies from their working equations.
These small size-inconsistency errors may be due in part to the size-extensivity errors introduced by truncating our theory. The fact that q-NOECCSD has a larger error than l-NOECCSD would align with the presence of disconnected terms in the q-NOECCSD equations that do not appear in l-NOECCSD. This suggests that taking the full NOECCSD equations (5-NOECCSD for the Li atom and 6-NOECCSD for the dimer) could remove the errors, although this would be prohibitively expensive in our current version of the code.

%{\color{blue} A similar observation was made by Hanauer and K\"ohn, where only certain ways of removing redundancies can lead to a size-consistent ic-MRCC formulation.} \cite{hanauer_pilot_2011}

%{\color{red} can we say something about other methods here? which others are properly size consistent and which aren't?}
%{\color{blue} Note point in 5.4. Kohn has also shown that only certain ways of handling redundancies lead to size consistent energies.}
% These terms can be corrected for by the addition of single excitation terms in l-CCSD and q-NOECCSD, therefore restoring size consistency.

% As dicussed by Mukherjee and Lindgren [REF], the excitation amplitudes for a single excitation ($t_{q}^{p}$) and its corresponding double excitation ($t_{qu}^{pu}$) with a spectator are related by $t_{q}^{p} + t_{qu}^{pu} = k$, where $k$ is a constant. 

% In the dissociated dimer the spin-orbital-based CCD still neglects singles, but the orbital-based NOECCD includes simultaneous orbital relaxation on both atoms, leading to the smaller size consistency error of 25-30$\mathrm{\mu E_{H}}$.
% In NOECCSD, the orbital relaxation on the single atom is captured, and size inconsistencies are reduced below the micro-Hartree level. 
% At this level of theory, the energies of both open-shell singlet and triplet states agree to milli-Hartree accuracy. TODO NICK: Should probably talk about this

\subsection{Orbital invariance}

The NOECC method proposed is invariant to core-core and virtual-virtual rotations in the same way as standard coupled-cluster theory. 
The reference CSF is not in general invariant to active-active rotations. 
This is in contrast to a CASSCF reference function, used in ic-MRCC methods, which is invariant to active-active rotations\cite{kong_orbital_2009}.
For any given excitation order, the NOE cluster operator $\hat T$ consists of all possible excitations from the set of core and active orbitals to the set of active and virtual orbitals. The NOE wave operator $\{ e^{\hat{T} } \}$ is therefore invariant to orbital rotations within each of these sets. If a certain subset of these excitations were excluded, the wave operator ceases to be orbital invariant. Our proposed NOECC method includes all excitations of a given type and therefore retains all orbital invariance properties of the reference CSF. 

\subsection{Compatibility with SRCC}

The state-specific NOECC method we have proposed is a single-reference method in the sense that we correlate a single reference CSF representative of the electronic eigenstate under consideration. In general, the CSF is a specific linear combination of many determinants. In the case that the CSF is a single closed-shell determinant, NOECC reduces identically to conventional spin-free coupled-cluster theory,  making it seamlessly compatible with SRCC methods.
Many of the problems associated with combining results from MRCC methods with SRCC methods stem from the use of a CASSCF reference, where it is often challenging to keep the reference space consistent across different regions of the potential energy surface. In the context of state-specific NOECC, this problem becomes one of choosing an appropriate CSF reference. 
Some regions of the potential energy surface will not be well represented by a single CSF\cite{martidafcik2024spincoupled} and there the state-specific approach is expected to be less accurate.
%Our NOECC method targets specific spin-states and can be safely combined with SRCC energies provided that a suitable CSF reference is

\subsection{Projected Schr\"odinger equation}
A proper residual equation\cite{kong_connection_2008, lyakh_multireference_2011, kohn_state-specific_2012} is defined as one that is equivalent to solving a projected Schr\"odinger equation. 
In general, the amplitude equations for methods using the JM \textit{ans\"atze}, where each reference determinant has its own wave operator, do not correspond to a projected
Schr\"odinger equation (a notable exception being Hanrath's MRexpT\cite{hanrath_exponential_2005, hanrath_initial_2006} method). The NOECC
approach uses an internally-contracted \textit{ansatz} with a complete first-order interacting space and therefore does satisfy the condition of having a proper residual equation.\\

\section{Conclusion}
We have developed a novel formulation of the single-reference normal-ordered exponential coupled-cluster method to correlate multi-determinant states, NOECC. The \textit{ansatz} is rigorously spin-adapted and recovers the dynamic correlation and orbital relaxation of an arbitrary configuration state function, without spin contamination. Both high- and low-spin states of an atom or molecule can be correlated.
Our working equations are derived from a reformulation of the unlinked coupled-cluster equations, which we prove are equivalent to solving fully connected equations. The method formally terminates at $4+\min(n,2N_{r})$ in the cluster amplitudes, where $n$ is the number of open-shell orbitals and $N_r$ is the maximum excitation rank of the cluster operator, at which order the theory is fully size extensive. In this way, we circumvent the requirement for the inverse of the normal ordered exponential, for which the closed form is not known. We have developed code to automatically generate spin-adapted equations in a truncated form, while keeping the size extensivity errors as small as possible.
The NOECCSD method truncated at second-order has been examined numerically using a set of small atoms and molecules, with encouraging results. Our energies for doublet systems are comparable to those found by Szalay and Gauss using a spin-restricted formalism and the singlet-triplet energy splitting are shown to be in excellent agreement with FCI for the 1s2s configuration of beryllium and within 10 kJ/mol of experiment for the oxygen molecule. Numerical tests of size consistency reveal that, while the method is not
rigorously size consistent, size-inconsistency errors are on the order of $\mu\mathrm{E_H}$ for the cases tested. In common with many MRCC methods, the NOECC wavefunction contains spectator excitations that lead to a set of redundant amplitudes in the residual equations. Although this leads to an infinite family of solutions, we find that different amplitude solutions yield
exactly the same energies.
NOECC is a single-reference method in the sense that coefficients of the multi-determinant reference state are not relaxed. Since our formulation does not rely on the absence of active-to-active excitations in the cluster operator, it can in principle therefore be used to correlate single CSFs, or CASSCF, RASSCF\cite{malmqvist_restricted_1990} or GASSCF\cite{ma_generalized_2011} references to recover the dynamic correlation of highly multi-reference states, without spin contamination. While for cases such as
bond dissociation, or near degeneracies, a fully multi-reference approach is more appropriate, NOECC is highly suited for correlating specific spin states, such as those in organic radicals and high- or low-spin transition metal spin states. \\

\section{Acknowledgements}
The authors would like to acknowledge Professor Andreas K\"ohn and Professor Debashis Mukherjee for providing insightful comments on this manuscript.\\

\appendix

\section{Connectedness of the residual equation} \label{A-Connectedness}
We demonstrate here that our working equations (Equation \ref{Theory:AmplitudeEq}) are equivalent to a connected form of the residual equations.
We also show that truncating our equations does introduce disconnected terms, but that size extensivity can be restored if the equations are taken to the full (finite) order at which they terminate.
The proof uses a derivation found in Lindgren's paper\cite{lindgren_coupled-cluster_1978} that extracts the connected terms from the product of the Hamiltonian and the normal ordered exponential form of the wave operator:
\begin{align}
    \hat{H} \{ e^{\hat{T}} \} &=  \{\hat{H} \{e^{\hat{T}}\}\} + \{\wick{\c{\hat{H}} \{\c{e}^{\hat{T}}\}}\}
\label{eqn:muk-identity-derivation}
\end{align}
The contracted terms can be separated into products of the connected parts and the remaining cluster operators:
\begin{align}
    \left\{\wick{\c{\hat{H}} \c{e}^{\hat{T}}}\right\} &= \sum_{n=1}^{\infty}\frac{1}{n!}\{\wick{\c{\hat{H}}\,\{\c{\hat{T}}^{n} \}}\}\\
    &=\sum_{n=1}^{\infty}\frac{1}{n!}\sum_{k=1}^{n}{n\choose k}\{(\hat{H}\,\hat{T}^{k})_{C}\,\}\{\hat{T}^{n-k}\}\} \}\\
    &=\sum_{(n-k)=0}^{\infty}\sum_{k=1}^{\infty}\frac{1}{k!}\frac{1}{(n-k)!}\{(\hat{H}\,\{\hat{T}^{k}\})_{C}\,\{\hat{T}^{n-k}\} \}\\
    &=\sum_{k=1}^{\infty}\frac{1}{k!}\{(\hat{H}\,\{\hat{T}^{k}\})_{C}\,\{e^{\hat{T}}\} \}
\label{eqn:muk-identity-contractions}
\end{align}
The first term of equation \ref{eqn:muk-identity-derivation} corresponds to extending the sum to $k=0$, leading to the classic result that\cite{mukherjee_correlation_1975, lindgren_coupled-cluster_1978}
\begin{align}
    \hat{H} \{ e^{\hat{T}} \} =\sum_{k=0}^{\infty}\frac{1}{k!}\{(\hat{H}\,\{\hat{T}^{k}\})_{C}\,\{e^{\hat{T}}\} \} =  \{ \{ e^{\hat{T}} \} (\hat{H} \{e^{\hat{T}}\})_{C}\}
\label{eqn:muk-identity}
\end{align}
where the subscript $C$ denotes connected terms.
We have also used the the fact that, within a normal ordered product, operators consisting of even numbers of creation and annihilation operators commute.

\subsection{Size extensivity in the closed shell case}
Our residual equation is
\begin{align}
    R_{I}&=\braket{ \Phi_{I} |\hat{H} \{ e^{\hat{T}} \} | \Phi_{0} }- \braket{ \Phi_{I} |\{ e^{\hat{T}} \} |{\Phi_{0}}}\braket{\Phi_{0}|\hat{H} \{ e^{\hat{T}} \} | \Phi_{0} } \\
    &= \braket{ \Phi_{I} |\{ \{ e^{\hat{T}} \} (\hat{H} \{e^{\hat{T}}\})_{C}\} | \Phi_{0} }- \braket{ \Phi_{I} |\{ e^{\hat{T}} \} |{\Phi_{0}}}\braket{\Phi_{0}|\hat{H} \{ e^{\hat{T}} \} | \Phi_{0} } \\
    &= \braket{ \Phi_{I} |\{ e^{\hat{T}} \} (\hat{H} \{e^{\hat{T}}\})_{C}| \Phi_{0} }-\braket{ \Phi_{I} |\wick{\c{\{ e^{\hat{T}} \}} \c{(\hat{H} \{e^{\hat{T}}\})_{C}}}| \Phi_{0} }\\
    &\quad- \braket{ \Phi_{I} |\{ e^{\hat{T}} \} |{\Phi_{0}}}\braket{\Phi_{0}|\hat{H} \{ e^{\hat{T}} \} | \Phi_{0} }
\end{align}
In the closed shell case, amplitudes cannot be contracted with from the right so the middle term is zero.
The disconnected terms in $\braket{\Phi_{0}|\hat{H} \{ e^{\hat{T}} \} | \Phi_{0} }$ will also vanish.
We then have
\begin{align}
    R_{I}&= \braket{ \Phi_{I} |\{ e^{\hat{T}} \} (\hat{H} \{e^{\hat{T}}\})_{C}| \Phi_{0} }- \braket{ \Phi_{I} |\{ e^{\hat{T}} \} |{\Phi_{0}}}\braket{\Phi_{0}|(\hat{H} \{e^{\hat{T}}\})_{C} | \Phi_{0} }\\
    &= \sum_{K\neq0}\braket{ \Phi_{I} |\{ e^{\hat{T}} \}|\Phi_{K}}\braket{\Phi_{K}| (\hat{H} \{e^{\hat{T}}\})_{C}| \Phi_{0} }\\
    &= \braket{ \Phi_{I}| (\hat{H} \{e^{\hat{T}}\})_{C}| \Phi_{0}} + \sum_{K\neq0}\braket{ \Phi_{I} |\{ e^{\hat{T}} -1\}|\Phi_{K}}\braket{\Phi_{K}| (\hat{H} \{e^{\hat{T}}\})_{C}| \Phi_{0}}
\end{align}
If $\ket{\Phi_{I}}$ is to be reached from $\ket{\Phi_{K}}$ using only excitations and $\ket{\Phi_{K}}\neq\ket{\Phi_{0}}$, then $\ket{\Phi_{K}}$ must be in the manifold spanned by the states $\ket{\Phi_{I}}$.
If $\ket{\Phi_{K}}$ is an excited state that can only be reached from $\ket{\Phi_{0}}$, then $\braket{\Phi_{K}| (\hat{H} \{e^{\hat{T}}\})_{C}| \Phi_{0}}=R_{K}$.
The disconnected terms vanish at convergence because they are comprised of smaller connected terms that are already solved in our equations.
Solving our (not fully connected) equation is equivalent to solving the connected equation
\begin{equation}
    0=R_{I}=\braket{ \Phi_{I}| (\hat{H} \{e^{\hat{T}}\})_{C}| \Phi_{0}}
\end{equation}

\subsubsection{The truncated residual equation}
Expanding our equation (including disconnected terms) for $R_{I}$ at each order in $\hat{T}$:
\begin{align}
    R_{I}^{(n)}&=\frac{1}{n!}\braket{ \Phi_{I} |\hat{H} \{ \hat{T}^{n}\} | \Phi_{0} }\\
    &\quad -\sum_{a=0}^{n}\frac{1}{a!}\braket{ \Phi_{I} |\{\hat{T}^{a}\} |{\Phi_{0}}}\frac{1}{(n-a)!}\braket{\Phi_{0}|\hat{H} \{ \hat{T}^{(n-a)}\} | \Phi_{0} }\\
    &=\frac{1}{n!}\sum_{a=0}^{n}{{n}\choose{a}}\braket{ \Phi_{I} |\{\hat{T}^{a}\}\left(\hat{H} \{ \hat{T}^{(n-a)}\}\right)_{C} | \Phi_{0} }\\
    &\quad -\frac{1}{n!}\sum_{a=0}^{n}{{n}\choose{a}}\braket{ \Phi_{I} |\{\hat{T}^{a}\} |{\Phi_{0}}}\braket{\Phi_{0}|\hat{H} \{ \hat{T}^{(n-a)}\} | \Phi_{0} }\\
    &=\frac{1}{n!}\sum_{a=0}^{n}{{n}\choose{a}}\sum_{K\neq0}\braket{ \Phi_{I} |\{\hat{T}^{a}\} |{\Phi_{K}}}\braket{\Phi_{K}|\left(\hat{H} \{ \hat{T}^{(n-a)}\}\right)_{C} | \Phi_{0} }
\end{align}
Our residual including terms up to order $N$ is then
\begin{align}
    R_{I}^{[N]}=&\sum_{n=0}^{N}\frac{1}{n!}\sum_{a=0}^{n}{{n} \choose{a}}\sum_{K\neq0}\braket{ \Phi_{I} |\{\hat{T}^{a}\} |{\Phi_{K}}}\braket{\Phi_{K}|\left(\hat{H} \{ \hat{T}^{(n-a)}\}\right)_{C} | \Phi_{0} }\\
    =&\sum_{n=0}^{N}\frac{1}{n!}\big(\braket{\Phi_{I}|\left(\hat{H} \{ \hat{T}^{n}\}\right)_{C} | \Phi_{0} }\\
    &\quad +\sum_{a=1}^{n}{{n}\choose{a}}\sum_{K\neq0}\braket{ \Phi_{I} |\{\hat{T}^{a}\} |{\Phi_{K}}}\braket{\Phi_{K}|\left(\hat{H} \{ \hat{T}^{(n-a)}\}\right)_{C} | \Phi_{0} }\big)
\end{align}
The disconnected terms on the right are no longer zero in general, as we are not solving the lower-truncated residual $R_{K}^{[N-a]}$, and our method is not size-extensive.

We do however avoid the spurious higher-order disconnected terms that would arise if each exponential were independently truncated at order $N$:
\begin{align}
    \Tilde{R}_{I}^{[N]}&=\sum_{a=0}^{N}\frac{1}{a!}\braket{ \Phi_{I} |\hat{H} \{ \hat{T}^{a}\} | \Phi_{0} }\\
    &\quad -\sum_{a=0}^{N}\frac{1}{a!}\braket{ \Phi_{I} |\{\hat{T}^{a}\} |{\Phi_{0}}}\sum_{b=0}^{N}\frac{1}{b!}\braket{\Phi_{0}|\hat{H} \{ \hat{T}^{b}\} | \Phi_{0} }
\end{align}

This does not affect the fact that the equations for closed shells terminate at $4$th order in the the amplitudes.
If our equations are truncated at $4$th order, we do recover the full size extensive formulation of the theory without truncation.
In the smaller connected components of the disconnected terms where $\ket{\Phi_{I}}$ can be reached from $\ket{\Phi_{K}}$ by application of $\{\hat{T}^{a}\}$, $R_{K}^{[N]}$ only has terms up to order $N-a$, so $R_{K}^{[N]}=R_{K}^{[N-a]}$ and our equation is equivalent to the connected form.

\subsection{The general case}
We first use the fact that the product of Hamiltonian and normal ordered wave operator $\hat{H} \{ e^{\hat{T}} \}$ can be expressed as the product of $\{ e^{\hat{T}} \}$ with a fully connected operator:
%with a sum of fully connected terms $\tilde{H}_{p}$
% \begin{equation}
%     \hat{H} \{ e^{\hat{T}} \} = \{ e^{\hat{T}} \} \sum_{p=0}^{\infty} (-1)^{p} \tilde{H}_{p}
% \end{equation}
\begin{equation}
    \hat{H} \{ e^{\hat{T}} \} = \{ e^{\hat{T}} \}\hat{\Lambda}
\end{equation}
where $\hat{\Lambda}$ is a sum of fully connected terms.
This result was first published by Mukherjee\cite{sen_inclusion_2018, chakravarti_reappraisal_2021}, and we will reproduce the proof here. 
It is convenient to define the fully connected term $\tilde{H}_{0} = (\hat{H} \{e^{\hat{T}}\})_{C}$ and the modified exponential $\{ \tilde{e}^{\hat{T}} \}=\{ e^{\hat{T}}-1 \}$.

Repeatedly applying Wick's theorem to equation \ref{eqn:muk-identity} yields the relation obtained by Chakravarti, Sen, and Mukherjee\cite{sen_inclusion_2018, chakravarti_reappraisal_2021}, that
\begin{align}
    \hat{H} \{ e^{\hat{T}} \} = \{ \{ e^{\hat{T}} \} \tilde{H}_{0}\} =\{ e^{\hat{T}} \} \hat{\Lambda}
\end{align}
where $\hat{\Lambda}$ is the sum of connected terms defined by
\begin{align}
    \hat{\Lambda} &= \{\tilde{H}_{0}\} - \{ (\{ \tilde{e}^{\hat{T}}\}\tilde{H}_{0})_{C} \} + \{ (\{ \tilde{e}^{\hat{T}}\}(\{ \tilde{e}^{\hat{T}}\}\tilde{H}_{0})_{C})_{C} \} \\
    &\quad - \{ (\{ \tilde{e}^{\hat{T}}\}(\{ \tilde{e}^{\hat{T}}\}(\{ \tilde{e}^{\hat{T}}\}\tilde{H}_{0})_{C})_{C})_{C} \} + \dots
\end{align}

Using this result, the residual equation can be written
\begin{align}
    R_{I}&=\braket{ \Phi_{I} |\hat{H} \{ e^{\hat{T}} \} | \Phi_{0} }- \braket{ \Phi_{I} |\{ e^{\hat{T}} \} |{\Phi_{0}}}\braket{\Phi_{0}|\hat{H} \{ e^{\hat{T}} \} | \Phi_{0} } \\
    &=\braket{ \Phi_{I} |\{ e^{\hat{T}} \} \hat{\Lambda}| \Phi_{0} }- \braket{ \Phi_{I} |\{ e^{\hat{T}} \} |{\Phi_{0}}}\braket{\Phi_{0}|\{ e^{\hat{T}} \} \hat{\Lambda} | \Phi_{0} }\\
    &= \sum_{K \neq 0} \Big( \braket{ \Phi_{I} | \{ e^{\hat{T}} \} | \Phi_{K}} - \braket{ \Phi_{I} |  \{ e^{\hat{T}} \} | \Phi_{0} }\braket{\Phi_{0}|\{ e^{\hat{T}} \} | \Phi_{K} }  \Big)\braket{ \Phi_{K} | \hat{\Lambda} | \Phi_{0} }\\
    &= \braket{ \Phi_{I} | \hat{\Lambda} | \Phi_{0} } \\
    &\quad +\sum_{K \neq 0} \Big( \braket{ \Phi_{I} | \{ \tilde{e}^{\hat{T}} \} | \Phi_{K}} - \braket{ \Phi_{I} |  \{ \tilde{e}^{\hat{T}} \} | \Phi_{0} }\braket{\Phi_{0}|\{ \tilde{e}^{\hat{T}} \} | \Phi_{K} }  \Big)\braket{ \Phi_{K} | \hat{\Lambda} | \Phi_{0} }
\label{Eq:Appendix-Amplitude-Intermediate-correct}
\end{align}
where the $K=0$ terms cancel due to the intermediate normalisation condition $\braket{\Phi_{0}|\{ e^{\hat{T}} \} | \Phi_{0} }=1$.

Assuming intermediate normalisation also gives that $\braket{\Phi_{0}|\{ \tilde{e}^{\hat{T}} \} | \Phi_{K} } =0$, so we can write
\begin{equation}
    R_{I}= \braket{ \Phi_{I} | \hat{\Lambda} | \Phi_{0} }+\sum_{K \neq 0} \braket{ \Phi_{I} | \{ \tilde{e}^{\hat{T}} \} | \Phi_{K}}\braket{ \Phi_{K} | \hat{\Lambda} | \Phi_{0} }
\end{equation}
As for the closed shell case, if $\ket{\Phi_{K}}$ cannot be reached from another excited state, then we have $\braket{ \Phi_{K} | \hat{\Lambda} | \Phi_{0} }=R_{K}$,
which vanishes at convergence.

Solving the residual Eq \ref{Theory:AmplitudeEq} is therefore equivalent to solving the fully connected equation
\begin{equation}
    0=R_{I}=\braket{ \Phi_{I}| \hat{\Lambda}| \Phi_{0}}
\end{equation}

The contribution of order $\hat{T}^{n}$ to the residual is
\begin{equation}
    R_{I}^{\left(n\right)} = \sum_{K\neq 0}\sum_{c=0}^{n}\frac{1}{(n-c)!}\braket{ \Phi_{I} | \{T^{(n-c)}\} | \Phi_{K} }\braket{ \Phi_{K} | \Lambda^{(c)} | \Phi_{0} }
\end{equation}

where $\Lambda^{(c)}$ is the sum of (connected) terms in $\Lambda$ of order $c$ with respect to $\hat{T}$.

The residual truncated at order $N$ is then
\begin{align}
    R_{I}^{[N]}&=\sum_{n=0}^{N}R_{I}^{(n)}=\sum_{K\neq 0}\sum_{c=0}^{N}\sum_{a=0}^{N-c}\frac{1}{a!}\braket{ \Phi_{I} | \{T^{a}\} | \Phi_{K} }\braket{ \Phi_{K} | \Lambda^{(c)} | \Phi_{0} }
\end{align}

where the relabelling $a=(n-c)$ has been made inside the sum.
Equivalently,
\begin{align}
    R_{I}^{[N]}&=\sum_{K\neq 0}\sum_{a=0}^{N}\frac{1}{a!}\braket{ \Phi_{I} | \{T^{a}\} | \Phi_{K} }\sum_{c=0}^{N-a}\braket{ \Phi_{K} | \Lambda^{(c)} | \Phi_{0} }\\
    &=\sum_{K\neq 0}\sum_{a=0}^{N}\frac{1}{a!}\braket{ \Phi_{I} | \{T^{a}\} | \Phi_{K} }\braket{ \Phi_{K} | \Lambda^{[N-a]} | \Phi_{0} }
\end{align}

where $\Lambda^{[N-a]}$ indicates the sum of terms in $\Lambda$ up to order $N-a$ in $\hat{T}$.
Since we have not solved the connected equations $\braket{ \Phi_{K} | \Lambda^{[N-a]} | \Phi_{0} }=0$ truncated to lower order, the truncation of our equations has once again introduced a size-extensivity error.

%%%REFERENCES%%%
\bibliography{bibliography} %You need to replace "rsc" on this line with the name of your .bib file
\bibliographystyle{rsc} %the RSC's .bst file

\end{document}